\documentclass[%
 reprint,
 amsmath,amssymb,
 aps,
]{revtex4-2}

\usepackage{graphicx}
\usepackage{dcolumn}
\usepackage{bm}
\usepackage{float}
\usepackage[many]{tcolorbox}
\usepackage{lipsum}
\usepackage{amsmath,cancel}

\usepackage{shuffle}
\usepackage{graphbox}
\usepackage{environ}
\usepackage{physics}
\usepackage{tabularx}
\graphicspath{figs/}

\begin{document}

\title{All-Loop Four-Point Aharony-Bergman-Jafferis-Maldacena Amplitudes from Dimensional Reduction of the Amplituhedron}

\author{Song He$^{1,2}$\footnote{songhe@itp.ac.cn}, Chia-Kai Kuo$^{3}$\footnote{chiakaikuo@gmail.com}, Zhenjie Li$^{1,4}$\footnote{lizhenjie@itp.ac.cn}, Yao-Qi Zhang$^{1,4}$\footnote{zhangyaoqi@itp.ac.cn}}
\affiliation{
$^{1}$CAS Key Laboratory of Theoretical Physics, Institute of Theoretical Physics, Chinese Academy of Sciences, Beijing 100190, China \\
$^{2}$School of Fundamental Physics and Mathematical Sciences, Hangzhou Institute for Advanced Study;\\
International Centre for Theoretical Physics Asia-Pacific, Beijing/Hangzhou, China\\
$^{3}$ Department of Physics and Center for Theoretical Physics, National Taiwan University, Taipei 10617, Taiwan\\
$^{4}$School of Physical Sciences, University of Chinese Academy of Sciences, No.19A Yuquan Road, Beijing 100049, China
}\date{\today}

\begin{abstract}
We define a new geometry obtained from the all-loop amplituhedron in ${\cal N}=4$ SYM by reducing its four-dimensional external and loop momenta to three dimensions. Focusing on the simplest four-point case, we provide strong evidence that the canonical form of this ``reduced amplituhedron" gives the all-loop integrand of the ABJM four-point amplitude. In addition to various all-loop cuts manifested by the geometry, we present explicitly new results for the integrand up to five loops, which are much simpler than results in ${\cal N}=4$ SYM. One of the reasons for such all-loop simplifications is that only a very small fraction of the so-called negative geometries survive the dimensional reduction, which corresponds to bipartite graphs. Our results suggest an unexpected relation between four-point amplitudes in these two theories.
\end{abstract}
\maketitle


\section{Introduction}
The amplituhedron in planar ${\cal N}=4$ SYM ~\cite{Arkani-Hamed:2013jha, Arkani-Hamed:2013kca,Arkani-Hamed:2017vfh} is arguably one of the most surprising mathematical structures of scattering amplitudes we have seen, where basic principles like locality and unitarity seem to have originated from the underlying geometric picture. All-loop integrands and tree amplitudes are given by the canonical form, which has logarithmic singularities only on boundaries of the amplituhedron~\cite{Arkani-Hamed:2012zlh, Arkani-Hamed:2017tmz}. As a beautiful mathematical object with remarkable physical properties, the amplituhedron has been extensively studied both at tree and loop level ({\it c.f.}~\cite{Lam:2014jda,Karp_2017,Galashin_2020, Arkani-Hamed:2013kca,Franco:2014csa,Galloni:2016iuj,Bai:2014cna,Bai:2015qoa, Rao:2018uta,Ferro:2018vpf}), and in particular it can be used to make all-loop predictions about cuts of the integrand~\cite{Arkani-Hamed:2018rsk,Langer:2019iuo}, which seem impossible otherwise. On the other hand, even for the four-point ($n=4$) $L$-loop amplituhedron, the geometry becomes more complicated as $L$ increases, and an explicit computation for $L \geq 4$ becomes rather difficult (though $n=4$ integrand has been known to $L=10$~\cite{Bern:2007ct,Bourjaily:2011hi,Bourjaily:2015bpz,Bourjaily:2016evz}). Moreover, despite various interesting ideas extending geometries beyond planar ${\cal N}=4$ SYM~\cite{Bern:2015ple, Arkani-Hamed:2017fdk, Arkani-Hamed:2018bjr, Arkani-Hamed:2017mur,Salvatori:2018aha, Arkani-Hamed:2019vag, Arkani-Hamed:2019mrd, Arkani-Hamed:2019plo, He:2018okq, Damgaard:2019ztj, Damgaard:2020eox}, 
an example of an all-loop amplituhedron in any other theory has yet to be found. 

By dimensionally reducing external and loop (region) momenta of the amplituhedron, we obtain a reduced amplituhedron with rich structures, but the computation of canonical forms becomes greatly simplified, at least for the $n=4$ case, which is a $3L$-dimensional geometry in the space of $L$ loop variables. Surprisingly, we find very strong evidence that this simplified $n=4$ geometry may be the long-sought-after all-loop amplituhedron for four-point amplitudes in ${\cal N}=6$ Aharony-Bergman-Jafferis-Maldacena (ABJM) theory~\cite{Aharony:2008ug}. In spite of an extensive literature on ABJM amplitudes at tree and one-loop level ({\it c.f.}~\cite{Huang:2010qy,Gang:2010gy, Bianchi:2012cq, Brandhuber:2012wy, Bargheer:2012cp}), much less is known about multi-loop ABJM integrands beyond $L=2$~\cite{Chen:2011vv,Bianchi:2011dg} (the only data available is a conjecture for $n=4$, $L=3$ in~\cite{Bianchi:2014iia}); even at tree-level, the amplituhedron in momentum space has only been proposed recently without obvious analog in momentum-twistor space yet~\cite{Huang:2021jlh,He:2021llb}. In this letter, we will not only show that the canonical forms of this $n=4$ reduced amplituhedron manifest various highly non-trivial all-loop cuts of ABJM amplitudes, but also push the frontier significantly by presenting compact expressions for ABJM integrands up to $L=5$. 

In ${\cal N}=4$ SYM, it is beneficial to decompose the $n=4$ amplituhedron into building blocks called {\it negative geometries}~\cite{Arkani-Hamed:2021iya}, and at each loop, non-trivial negative geometries combine to give the integrand for an infrared-finite observable closely related to the logarithm of amplitudes (or equivalently Wilson loops with a single insertion)~\cite{Alday:2011ga, Engelund:2011fg, Alday:2013ip, Henn:2019swt, Chicherin:2022bov, Chicherin:2022zxo}. The analogous decomposition of the reduced amplituhedron reveals enormous simplifications from $D=4$ to $D=3$: only a tiny fraction of negative geometries, namely those corresponding to {\it bipartite graphs}, contribute to the integrand, with very simple pole structures. This lies at the heart of all-loop simplifications when reducing the $n=4$ amplituhedron to $D=3$. 
\section{Dimensional reduction of the amplituhedron}

In this section, we first give the definition of the reduced amplituhedron for $n=4$, and then we outline a huge reduction of negative geometries in $D=3$ of the corresponding geometries in $D=4$~\cite{Arkani-Hamed:2021iya}. 
\subsection{Definition of reduced amplituhedron}
Recall that the $n$-point amplituhedron is defined in the space of $n$ momentum twistors~\cite{Hodges:2009hk},
$Z_a^I$ with $a=1,2,\dots, n$ for external kinematics, as well as $L$ lines in the twistor space, $(AB)^{I J}_i$ with $i=1, \dots, L$ for loop momenta; here $I,J=1,\dots, 4$ are $\operatorname{SL}(4)$ indices, and the simplest bosonic $\operatorname{SL}(4)$ invariant is defined as $\langle a b c d\rangle\equiv \epsilon_{I J K L}Z_a^I Z_b^J Z_c^K Z_d^L$ (and similarly for $\langle (AB)_i a b\rangle$ and $\langle (AB)_i (AB)_j\rangle$). In~\cite{Elvang:2014fja}, external kinematics in $D=3$ was defined by dimensionally reducing every external line, $Z_a Z_{a{+}1}$; in a completely analogous manner, here we also need to dimensionally reduce all loop variables $(AB)_i$, both of which are achieved by the so-called {\it symplectic conditions} on these lines:
\begin{equation}\label{sympletic}
{\bf \Omega}_{IJ} Z_a^I Z_{a{+}1}^J={\bf \Omega}_{IJ} A_i^I B_i^J=0\,, {\rm with}~{\bf \Omega}=\mqty(0 & \epsilon_{2\times 2} \\ \epsilon_{2\times2} & 0)
\end{equation}
for $a=1, 2, \dots, n$ and $i=1, \dots, L$, where the totally antisymmetric matrix is defined as $\epsilon_{2\times 2}=\mqty(0 & 1 \\ -1 & 0)$. 

We can define the reduced amplituhedron for any $n$ and $L$ by restricting the $D=4$ amplituhedron geometry on the subspace given by \eqref{sympletic}, as long as it has a non-vanishing support there. In this letter, we focus on the special case $n=4$ 
, and it is clear that we have a $3L$-dimensional geometry defined in the projected $(AB)_{i=1,\dots, L}$ space. An important subtlety is that $\langle 1234\rangle<0$ for real $Z$'s satisfying symplectic conditions, thus we need to flip the overall sign for the definition of the $D=4$ amplituhedron~\cite{Arkani-Hamed:2013jha}: we require $\langle AB 12\rangle, \langle AB 23\rangle, \langle AB 34\rangle, \langle AB 14\rangle<0$ and $\langle AB 13\rangle, \langle AB 24\rangle>0$, for any loop $(AB)$, as well as $\langle (A B)_i (AB)_j\rangle<0$, all on the support of \eqref{sympletic}.

A convenient parametrization is $(A B)_i=(Z_1+ x_i Z_2- w_i Z_4, y_i Z_2+ Z_3+z_i Z_4)$~\cite{Arkani-Hamed:2013kca}, and the symplectic condition on $(AB)_i$ becomes $x_i z_i+ y_i w_i-1=0$; the $n=4$ geometry is defined by ($x_{i,j}:=x_i- x_j$ {\it etc.})
\begin{align}\label{def1}
&\forall i: x_i, y_i, z_i, w_i>0, \quad x_i z_i+y_i w_i=1,\nonumber\\
&\forall i,j: x_{i,j} z_{i,j} + y_{i,j} w_{i,j}<0.
\end{align}
We denote this geometry as ${\cal A}_L$ with the canonical form $\Omega({\cal A}_L):=\Omega_L$, and our main claim is that $\Omega_L$ gives the $L$-loop planar integrand for four-point ABJM amplitudes (after stripping off the overall tree amplitude). 

\subsection{Negative geometries and their dimension reduction}\label{sec:2}
In~\cite{Arkani-Hamed:2021iya}, a nice rewriting for the $n=4$ amplituhedron~\cite{Arkani-Hamed:2013kca} was proposed, where it is decomposed into a sum of negative geometries given by ``mutual negativity" conditions, which trivially carries over to our ${\cal A}_L$ in $D=3$; each negative geometry is represented by a labelled graph with $L$ nodes and $E$ edges (edge $(i j)$ for $\langle (AB)_i (AB)_j\rangle>0$ since we reversed all signs, and no condition otherwise), with an overall sign factor $(-)^E$. We sum over all graphs with $L$ nodes without $2$-cycles, 
\begin{equation}
{\cal A}_L=\sum_g (-)^{E(g)} {\cal A}(g)
\end{equation} 
where ${\cal A}(g)$ is the (oriented) geometry for graph $g$. It suffices to consider all {\it connected} graphs, whose (signed) sum gives the geometry for the logarithm of amplitudes~\cite{Arkani-Hamed:2021iya}. Such a  decomposition is useful since each $A_g$ is simpler, whose canonical form is easier to compute. The form for $L=2,3$ reads:
\begin{equation}
    \centering
    \includegraphics[scale=0.9]{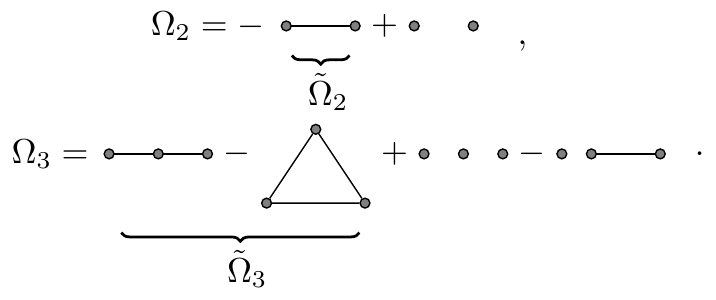}
\end{equation}
where the connected part, or $\log$ of the amplitude, is denoted as $\tilde{\Omega}_L$, {\it e.g.} $\tilde{\Omega}_2:=\Omega_2-\frac 1 2 \Omega_1^2$. Similarly, the connected part of $L=4$ is given by the sum of graphs with $6$ topologies (and so on for higher $L$),
\begin{equation}
    \includegraphics{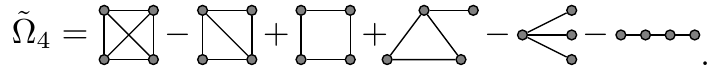}
\end{equation}


What is new in $D=3$ is that most of these geometries do not contribute at all: we find that remarkably, under dimensional reduction only those negative geometries with {\it bipartite} graphs survive in the decomposition. For example, for $\tilde{\Omega}_3$, the chain graph contributes but the triangle does not; while for $\tilde{\Omega}_4$, only the two kinds of tree graphs and the box contribute. This represents a major simplification as the fraction of bipartite graphs in all graphs tends to zero quickly as $L$ increases: for $L=2,\dots, 7$, the number of topologies for connected graphs are $1,2,6,21,112, 853$, but that of bipartite topologies decrease to $1,1,3, 5, 17, 44$, {\it e.g.} for $\tilde{\Omega}_5$, only $5$ topologies (out of $21$) survive the reduction.  Moreover, it turns out that one can compute the canonical form for geometries of bipartite graphs with relative ease, mainly due to their remarkably simple pole structures. The reduction and the computation of their forms will be explained in detail in~\footnote{See Supplemental Material for the detail of the reduction and the computation of their forms, which includes Ref.~\cite{stanley1973acyclic}.}.

\section{ABJM integrands from reduced amplituhedron}
Let us take a first look at $L=1$, where the geometry is defined as $x,y,z,w>0$ and $x z + y w=1$. In this special case, its canonical form is nothing but reducing the $D=4$ form, $\frac{d x}{x} \frac{d y}{y} \frac{d z}{z} \frac{d w}{w}$, onto the $D=3$ subspace: 
\begin{equation}
\Omega_1=\frac{d x}{x} \frac{d y}{y} \frac{d z}{z} \frac{d w}{w} \delta(x z + y w-1),
\end{equation}
and we can rewrite it in a covariant form:
\begin{equation}\Omega_1=\frac{d^3 AB \langle 1234\rangle^{3/2} (\langle AB13\rangle \langle AB24\rangle)^{1/2}}{\langle AB12\rangle \langle AB23\rangle \langle AB34\rangle \langle AB14\rangle },
\end{equation}
where the measure is
\begin{equation}
 d^3 AB{:=}\langle AB d^2 A\rangle \langle AB d^2 B\rangle \delta({\bf \Omega}_{IJ} A^I B^J).
\end{equation} 
Here the numerator, $(\langle AB13\rangle \langle AB24\rangle)^{1/2}{\propto}(x z+y w)^{1/2}=1$, turns out to be proportional to the famous $\epsilon$-numerator of the  dual conformal invariant box in $D=3$~\cite{Chen:2011vv}.  Thus the dimensional reduction of the one-loop box in ${\cal N}=4$ SYM gives the one-loop box with the $\epsilon$-numerator in ABJM, which confirms our claim at one loop.

\subsection{All-loop soft and vanishing cuts}
Now we provide strong evidence that $\Omega_L$ gives $L$-loop integrand for $L>1$. Let us first rewrite the inequalities \eqref{def1} by solving for variables $x_i=(1- y_i w_i)/z_i$, and we arrive at the following equivalent definition
\begin{align}\label{def2}
& w_i, y_i, z_i>0, w_i y_i<1,\\
&
d_{i,j}:=(w_i z_j-w_j z_i)(y_i z_j-y_j z_i)-z_{i,j}^2<0, \nonumber
\end{align}
for $i,j=1, \dots, L$. Before proceeding to explicit computations, we see that \eqref{def2} allows us to study some all-loop cuts in a simple way. An important cut of four-point ABJM amplitudes is the so-called soft cut, where we take $\langle AB 12\rangle=\langle AB 23\rangle=\langle AB 34\rangle=0$, or equivalently $y=z=w=0$ for any given loop
, and the result is the $(L-1)$-loop integrand. From geometry, with $y_i=z_i=w_i=0$ clearly $d_{i,j}<0$ is trivially satisfied for any $j\neq i$, thus the geometry reduces to the $(L-1)$-loop one: $\partial^{(3)}_{y_i=z_i=w_i=0} {\cal A}_L={\cal A}_{L-1}$. The soft cut is satisfied! 

Moreover, certain cuts are known to vanish due to the presence of vanishing odd-point amplitudes: by cutting $\langle (AB)_i 12\rangle=\langle (AB)_i (AB)_j\rangle=\langle (AB)_j 12\rangle=0$ (or change the last one to $\langle (AB)_j 34\rangle$), we isolate a three-point (or five-point, respectively) amplitude, which must vanish. These are equivalent to setting $w_i=d_{i,j}=w_j=0$ or $w_i=d_{i,j}=y_j=0$; in either case, $d_{i,j}=-z_{i,j}^2<0$ is trivially satisfied on the support of the other two conditions, and the residue vanishes as expected. These and other vanishing cuts are nicely guaranteed by the geometry.
We have checked our new results for $L=4,5$ thoroughly: in addition to various all-loop checks, we have computed unitarity cuts and checked that $\Omega_4$ and $\Omega_5$ satisfy the optical theorem: double-cuts are given by products of various lower-loop integrands.

\subsection{Explicit results up to five loops}
We present explicitly the canonical form up to five loops and leave the detailed derivation in the supplemental material. To save space, we introduce a shorthand notation $(AB)_i:=\ell_i$, and it turns out that the logarithm $\tilde{\Omega}_2$ is simply
\begin{equation}
\tilde{\Omega}_2=-2 \frac{d^3 \ell_1 d^3 \ell_2 \langle 1234\rangle^2}{\langle \ell_1 12\rangle \langle \ell_1 34 \rangle \langle \ell_1 \ell_2\rangle\langle \ell_2 2 3 \rangle\langle \ell_2 14\rangle} + (\ell_1 \leftrightarrow \ell_2).
\end{equation}
This is nothing but a double-triangle integrand where external region momenta correspond to $(12), (34)$ for $\ell_1$, and $(23), (14)$ for $\ell_2$, and vice versa. One can easily check that by adding back one-loop squared, we recover the well-known two-loop result~\cite{Chen:2011vv,Bianchi:2011dg}. 

One interesting feature of $\tilde{\Omega}_2$ is that for each term, $\ell_1$ contains only two poles, $\langle \ell_1 12\rangle \langle \ell_1 34 \rangle$ or $\langle \ell_1 23\rangle \langle \ell_1 14 \rangle$ (similarly for $\ell_2$). We denote these combinations and mutual conditions, which include all possible poles, as: 
\begin{equation*}
s_i:=\langle \ell_i 12 \rangle\langle \ell_i 34 \rangle,\quad  t_i:=\langle \ell_i 23\rangle \langle \ell_i 14 \rangle, \quad D_{i,j}:=-\langle \ell_i \ell_j\rangle.
\end{equation*}
We denote the $\epsilon$-numerator and $\ell$-independent factor as \begin{equation*}
 \epsilon_i:=(\langle \ell_i 13 \rangle \langle \ell_i 24 \rangle \langle 1234\rangle)^{1/2}, \quad c:= \langle 1234\rangle, 
\end{equation*} 
and also the integrand with measure $\prod_{i=1}^L d^3 \ell_i$ stripped off as $\underline{\tilde{\Omega}}_L$. For $L=1,2$, they read:
\begin{align}
&\underline{\Omega}_1=\frac{c \epsilon_1}{s_1 t_1},\nonumber\\
&\underline{\tilde{\Omega}}_2=\frac{2 c^2}{D_{1,2}}(\frac{1}{s_1 t_2}+ \frac{1}{t_1 s_2})=
    \includegraphics[align=c]{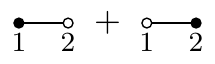},
\label{2loop}
\end{align}
where the $L=2$ case is represented by ``chain" graphs with $s,t$ pole structures represented by black and white coloring, respectively.

Now we are ready to move to $L=3$. Remarkably we find that $\tilde{\Omega}_3$ only receives contributions with pole structures of $3$ chains (each with two possible choices of $s$ and $t$), as represented by $6$ bipartite graphs. It reads
\begin{align}\label{3loop}
\underline{\tilde{\Omega}}_3=&
\includegraphics[align=c]{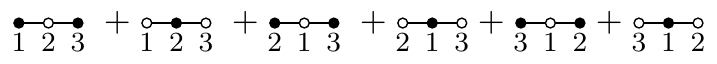}\nonumber\\
=&\frac{4 c^2 \epsilon_2 }{s_1 t_2 s_3 D_{1,2} D_{2,3}}+ (s\leftrightarrow t) +2~{\rm perms.\,,} 
\end{align}
where for each chain we have two terms with $s, t$ swapped, and similar to $L=1$ the $\epsilon$-numerator makes correct weight: $\underline{\tilde{\Omega}}_L$ has degree-$(-3)$ in each $\ell_i$. 
Each term is again a ladder integral with two triangles and a middle box (with $\epsilon$ numerator). Very non-trivially, when converting back to $\Omega_3$, it agrees with the conjecture from generalized unitarity~\cite{Bianchi:2014iia}; various independent checks are presented in the supplementary material. 
\begin{figure}[H]
    \centering
     \includegraphics{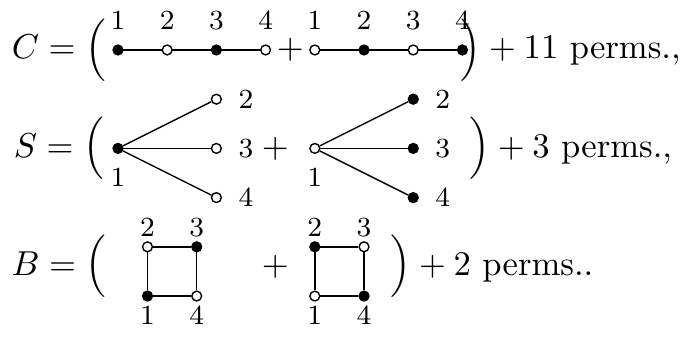}
    \caption{All bipartite graphs contributing to $\tilde{\Omega}_4$.}
    \label{4loops}
\end{figure}

After having familiarized ourselves with the notation, we present the $L=4$ result in a very compact form, and it turns out $\tilde{\Omega}_4$ only gets contributions from three topologies. We give all bipartite graphs in Fig.~\ref{4loops}. The first type consists of $12\times 2$ bipartite chain graphs
\begin{equation}
C=8 c^2
\frac{\epsilon_2 \epsilon_3}{D_{1,2} D_{2,3} D_{3,4} s_1  t_2 s_3 t_4} + (s\leftrightarrow t)+ 11~{\rm perms}.\,;
\end{equation}
then we have $4\times 2$ ``star" bipartite graphs:
\begin{equation}
S=
8 c^3
\frac{t_1} {D_{1,2} D_{1,3} D_{1,4} s_1 t_2 t_3 t_4}+ (s\leftrightarrow t) + 3~{\rm perms.}\,;
\end{equation}
finally we have $3$ box bipartite graphs:
\begin{align}\label{eq:L=4 box}
B=&\,4 \frac{4 \epsilon_1 \epsilon_2 \epsilon_3 \epsilon_4- 
c (\epsilon_1 \epsilon_3 N^t_{24} + \epsilon_2 \epsilon_4 N^s_{13})- 
c^2 N^{\rm cyc}_{1,2,3,4} }{D_{1,2} D_{2,3} D_{3,4} D_{4,1} s_1 t_2 s_3 t_4}\\\nonumber
&+(s \leftrightarrow t) + 2~{\rm perms.}\,,
\end{align}
where we define combinations similar to $s, t$ for two $\ell$s,
\begin{align}
  &N^s_{13}:=\langle \ell_1 12\rangle \langle \ell_3 34\rangle + \langle \ell_3 12\rangle \langle \ell_1 34\rangle\\\nonumber
  &N^t_{24}:=\langle \ell_2 14\rangle \langle \ell_4 23\rangle + \langle \ell_4 14\rangle \langle \ell_2 23\rangle\\\nonumber
  &N^{\rm cyc}_{i,j,k,l}:=\langle \ell_i 1 2\rangle  \langle \ell_j 3 4\rangle\langle \ell_k 1 2\rangle\langle \ell_l 3 4\rangle + {\rm cyc}(1,2,3,4),
\end{align}
where cyc$(1,2,3,4)$ indicates cyclic rotations of dual points $12\to 23 \to 34 \to 14$; $(s \leftrightarrow t)$ denotes the symmetrization in the pairs $(12, 34)$ and $(23, 14)$. 

The final result for $L=4$ reads 
\begin{equation}
\tilde{\underline\Omega}_4=-C-S+B,
\end{equation}
where the signs are given by $(-)^E$ with $E$ the number of edges. We compute these forms using a method whose details are given in the supplementary material: after writing down denominators according to bipartite graphs, we use an ansatz for each numerator which consists of all possible terms consistent with symmetries, and fix all parameters (with numerous cross checks) from various boundaries whose canonical forms can be computed directly. 

Finally, we compute the five-loop form $\tilde{\Omega}_5$, which consists of $5$ topologies: three tree graphs with $4$ edges, a box with an external line ($5$ edges), and one with $2$ nodes connected to $3$ nodes ($6$ edges). We have
\begin{equation}
    \includegraphics[scale=0.9]{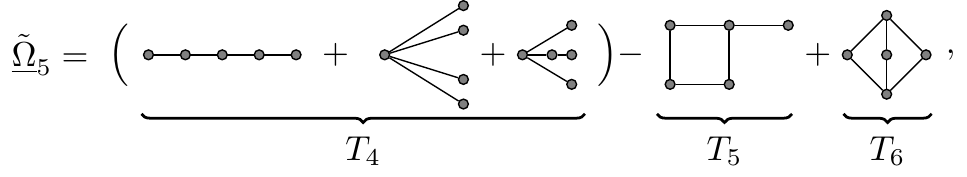}
\end{equation}
\noindent
where only graphs without labels or color are shown, and $T_m$ denotes the total contribution from graphs with $m$ edges. The contribution of all trees, $T_4$, takes similar forms as lower trees ({\it e.g.} $C,S$ for $L=4$).
We follow the same method for computing the remaining contributions: $60\times 2$ bipartite graphs for $T_5$ and $10\times 2$ for $T_6$, which are analogous to $B$ (especially $T_5$ takes a very similar form). The upshot is that the full five-loop result boils down to just these two new functions ($T_5$ and $T_6$) given as follows 

\begin{widetext}
\begin{align}
T_4&=
\includegraphics[align=c,scale=0.8]{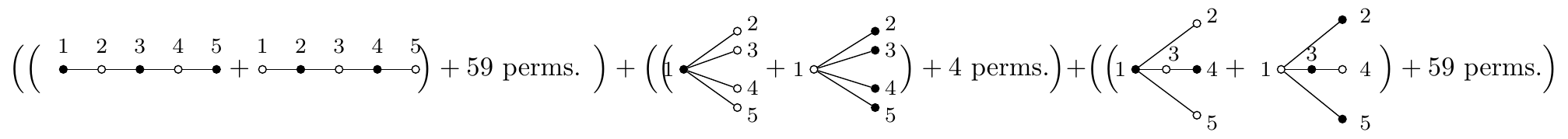},
\nonumber\\
&=\left(\frac{16c^3\epsilon_2\epsilon_4}{s_1t_2s_3t_4s_5 D_{1,2}D_{2,3}D_{3,4}D_{4,5}}+\left(s\leftrightarrow t\right)+59\,  \text{perms.}\right)+\left(\frac{16c^3\epsilon_1  s_1}{t_1s_2s_3s_4s_5D_{1,2}D_{1,3}D_{1,4}D_{1,5}}+\left(s\leftrightarrow t\right)+4\,  \text{perms.}\right)\nonumber\\
&+\left(\frac{16c^3\epsilon_3 t_1}{s_1t_2t_3s_4t_5D_{1,2}D_{1,3}D_{3,4}D_{1,5}}+\left(s\leftrightarrow t\right)+59\, \text{perms.}\right)\,,\nonumber\\
T_5&=
8c 
\frac{4 \epsilon_1 \epsilon_3 \epsilon_4 s_2- \epsilon_1 \epsilon_2 \epsilon_3 N_{24}^t-c (-\epsilon_1 t_2 N_{34}^t-\epsilon_3 t_2 N_{14}^t+\epsilon_4 s_2 N_{13}^s+\epsilon_2 N^{\rm cyc}_{1,2,3,4})}{s_1 t_2 s_3 t_4 s_5 D_{1,2} D_{2,3} D_{3,4} D_{4,1} D_{2,5}}+ (s\leftrightarrow t) + 59~ {\rm perms.}\,,\nonumber\\
T_6&=
\frac{4}{c
s_1 t_2 t_3 t_4 s_5 D_{1,2} D_{1,3} D_{1,4} D_{2,5} D_{3,5} D_{4,5}}\bigl( -8 \epsilon_1 \epsilon_2\epsilon_3 \epsilon_4 \epsilon_5 N_{15}^s+ c \epsilon_2 \epsilon_3 \epsilon_4 P_a \nonumber\\
&+ c [ \epsilon_1 \epsilon_2 \epsilon_3 P_b + (\ell_1\leftrightarrow \ell_5)+ \epsilon_1 \epsilon_2 \epsilon_5 P_c + {\rm cyc} (\ell_2, \ell_3, \ell_4)]\nonumber\\
&  + c^2 [\epsilon_1 P_d + (\ell_1\leftrightarrow \ell_5)]+c^2 [\epsilon_2 P_e + {\rm cyc} (\ell_2, \ell_3, \ell_4)] \bigr) + (s\leftrightarrow t) + 9~{\rm perms.};
\end{align}
\end{widetext}
where in $T_6$, we have polynomials $P_a, P_b, \dots, P_e$ with certain weights in $\ell_1, \dots, \ell_5$, and here we record their expressions: 
\begin{widetext}
\begin{align}
P_a&:=-20 s_1 s_5+16 t_1 t_5+(N_{15}^s)^2, \quad P_b:=6 s_5 N_{14}^s,
P_c:=N_{15}^s N_{34}^t-4 N^{\rm cyc}_{1,3,5,4},\\
P_d&:=-s_5 (N_{12}^s N_{34}^t + {\rm cyc}(\ell_2, \ell_3, \ell_4))+[2\langle \ell_5 1 2\rangle^2 \langle \ell_1 1 2\rangle \langle \ell_2 3 4\rangle \langle \ell_3 3 4\rangle \langle \ell_4 3 4\rangle + {\rm cyc}(1,2,3,4)]\nonumber\\
&+ 2t_5 [\langle \ell_1 1 4\rangle (\langle \ell_2 1 4\rangle \langle \ell_3 2 3 \rangle \langle \ell_4 2 3 \rangle + {\rm cyc} (\ell_2, \ell_3, \ell_4))  + (14 \leftrightarrow 23)],\\
P_e&:=2 s_1 s_5 (N_{34}^t-N_{34}^s)-4 t_1 t_5 N_{34}^t-[s_5 (\langle \ell_1 1 2\rangle^2 
\langle \ell_3 3 4\rangle \langle \ell_4 3 4\rangle 
+ (12 \leftrightarrow 34)) +(\ell_1 \leftrightarrow \ell_5)]\nonumber\\
&+N_{15}^s (\langle \ell_1 1 4\rangle \langle \ell_5 1 4\rangle \langle \ell_3 2 3 \rangle \langle \ell_4 2 3\rangle +(14 \leftrightarrow 23)).
\end{align}
\end{widetext}

\section{Conclusions and outlook}
In this letter we have discovered a surprising connection between four-point amplitudes in ${\cal N}=4$ SYM and ABJM: by dimensionally reducing from $D=4$ to $D=3$, the amplituhedron of the former becomes that of the latter, which we have checked explicitly to five loops and for various all-loop cuts. The reduced geometries exhibit remarkable structures and simplicity. 

One pressing question is the physical meaning of reduced amplituhedra for higher points: does it correspond to ABJM amplitudes 
or certain null polygonal Wilson loops~\cite{Bianchi:2011dg, Henn:2010ps}? On the other hand, our $n=4$ integrand clearly contains higher-point ones via unitarity, {\it e.g.} their single-cuts give forward-limit of six-point amplitudes~\cite{Huang:2013owa,Huang:2014xza} (see \cite{Caron-Huot:2012sos} for $L=2$). It would be fascinating to compute such higher-point forms at $L\geq 3$ and reveal possible geometries.

Last but not least, integrating the forms produces an interesting, finite observable in ABJM theory (analogous to that in ${\cal N}=4$ SYM~\cite{Arkani-Hamed:2021iya}). It is straightforward to do so for $L\leq 3$, and we expect to extract that cusp anomalous dimension of ABJM from it; we also expect that resummation for (some of) bipartite geometries would allow us to study their contributions non-perturbatively. We will report these results elsewhere.

\begin{acknowledgments}
We thank Nima Arkani-Hamed, Yu-tin Huang, Qinglin Yang for inspiring discussions, and especially Yu-tin Huang for clarifying the $\epsilon$ structure of ABJM.  The research of S. H. is supported in part by the National Natural Science Foundation of China under Grant No.11935013, 11947301, 12047502, 12047503. The research of C.-K. K. is supported by Taiwan Ministry of Science and Technology Grant No. 109-2112-M-002-020-MY3.
\end{acknowledgments}

\bibliographystyle{apsrev4-1}
\bibliography{bib}

\begin{thebibliography}{58}%
\makeatletter
\providecommand \@ifxundefined [1]{%
 \@ifx{#1\undefined}
}%
\providecommand \@ifnum [1]{%
 \ifnum #1\expandafter \@firstoftwo
 \else \expandafter \@secondoftwo
 \fi
}%
\providecommand \@ifx [1]{%
 \ifx #1\expandafter \@firstoftwo
 \else \expandafter \@secondoftwo
 \fi
}%
\providecommand \natexlab [1]{#1}%
\providecommand \enquote  [1]{``#1''}%
\providecommand \bibnamefont  [1]{#1}%
\providecommand \bibfnamefont [1]{#1}%
\providecommand \citenamefont [1]{#1}%
\providecommand \href@noop [0]{\@secondoftwo}%
\providecommand \href [0]{\begingroup \@sanitize@url \@href}%
\providecommand \@href[1]{\@@startlink{#1}\@@href}%
\providecommand \@@href[1]{\endgroup#1\@@endlink}%
\providecommand \@sanitize@url [0]{\catcode `\\12\catcode `\$12\catcode
  `\&12\catcode `\#12\catcode `\^12\catcode `\_12\catcode `\%12\relax}%
\providecommand \@@startlink[1]{}%
\providecommand \@@endlink[0]{}%
\providecommand \url  [0]{\begingroup\@sanitize@url \@url }%
\providecommand \@url [1]{\endgroup\@href {#1}{\urlprefix }}%
\providecommand \urlprefix  [0]{URL }%
\providecommand \Eprint [0]{\href }%
\providecommand \doibase [0]{http://dx.doi.org/}%
\providecommand \selectlanguage [0]{\@gobble}%
\providecommand \bibinfo  [0]{\@secondoftwo}%
\providecommand \bibfield  [0]{\@secondoftwo}%
\providecommand \translation [1]{[#1]}%
\providecommand \BibitemOpen [0]{}%
\providecommand \bibitemStop [0]{}%
\providecommand \bibitemNoStop [0]{.\EOS\space}%
\providecommand \EOS [0]{\spacefactor3000\relax}%
\providecommand \BibitemShut  [1]{\csname bibitem#1\endcsname}%
\let\auto@bib@innerbib\@empty
\bibitem [{\citenamefont {Arkani-Hamed}\ and\ \citenamefont
  {Trnka}(2014{\natexlab{a}})}]{Arkani-Hamed:2013jha}%
  \BibitemOpen
  \bibfield  {author} {\bibinfo {author} {\bibfnamefont {N.}~\bibnamefont
  {Arkani-Hamed}}\ and\ \bibinfo {author} {\bibfnamefont {J.}~\bibnamefont
  {Trnka}},\ }\href {\doibase 10.1007/JHEP10(2014)030} {\bibfield  {journal}
  {\bibinfo  {journal} {JHEP}\ }\textbf {\bibinfo {volume} {10}},\ \bibinfo
  {pages} {030} (\bibinfo {year} {2014}{\natexlab{a}})},\ \Eprint
  {http://arxiv.org/abs/1312.2007} {arXiv:1312.2007 [hep-th]} \BibitemShut
  {NoStop}%
\bibitem [{\citenamefont {Arkani-Hamed}\ and\ \citenamefont
  {Trnka}(2014{\natexlab{b}})}]{Arkani-Hamed:2013kca}%
  \BibitemOpen
  \bibfield  {author} {\bibinfo {author} {\bibfnamefont {N.}~\bibnamefont
  {Arkani-Hamed}}\ and\ \bibinfo {author} {\bibfnamefont {J.}~\bibnamefont
  {Trnka}},\ }\href {\doibase 10.1007/JHEP12(2014)182} {\bibfield  {journal}
  {\bibinfo  {journal} {JHEP}\ }\textbf {\bibinfo {volume} {12}},\ \bibinfo
  {pages} {182} (\bibinfo {year} {2014}{\natexlab{b}})},\ \Eprint
  {http://arxiv.org/abs/1312.7878} {arXiv:1312.7878 [hep-th]} \BibitemShut
  {NoStop}%
\bibitem [{\citenamefont {Arkani-Hamed}\ \emph
  {et~al.}(2018{\natexlab{a}})\citenamefont {Arkani-Hamed}, \citenamefont
  {Thomas},\ and\ \citenamefont {Trnka}}]{Arkani-Hamed:2017vfh}%
  \BibitemOpen
  \bibfield  {author} {\bibinfo {author} {\bibfnamefont {N.}~\bibnamefont
  {Arkani-Hamed}}, \bibinfo {author} {\bibfnamefont {H.}~\bibnamefont
  {Thomas}}, \ and\ \bibinfo {author} {\bibfnamefont {J.}~\bibnamefont
  {Trnka}},\ }\href {\doibase 10.1007/JHEP01(2018)016} {\bibfield  {journal}
  {\bibinfo  {journal} {JHEP}\ }\textbf {\bibinfo {volume} {01}},\ \bibinfo
  {pages} {016} (\bibinfo {year} {2018}{\natexlab{a}})},\ \Eprint
  {http://arxiv.org/abs/1704.05069} {arXiv:1704.05069 [hep-th]} \BibitemShut
  {NoStop}%
\bibitem [{\citenamefont {Arkani-Hamed}\ \emph {et~al.}(2016)\citenamefont
  {Arkani-Hamed}, \citenamefont {Bourjaily}, \citenamefont {Cachazo},
  \citenamefont {Goncharov}, \citenamefont {Postnikov},\ and\ \citenamefont
  {Trnka}}]{Arkani-Hamed:2012zlh}%
  \BibitemOpen
  \bibfield  {author} {\bibinfo {author} {\bibfnamefont {N.}~\bibnamefont
  {Arkani-Hamed}}, \bibinfo {author} {\bibfnamefont {J.~L.}\ \bibnamefont
  {Bourjaily}}, \bibinfo {author} {\bibfnamefont {F.}~\bibnamefont {Cachazo}},
  \bibinfo {author} {\bibfnamefont {A.~B.}\ \bibnamefont {Goncharov}}, \bibinfo
  {author} {\bibfnamefont {A.}~\bibnamefont {Postnikov}}, \ and\ \bibinfo
  {author} {\bibfnamefont {J.}~\bibnamefont {Trnka}},\ }\href {\doibase
  10.1017/CBO9781316091548} {\emph {\bibinfo {title} {{Grassmannian Geometry of
  Scattering Amplitudes}}}}\ (\bibinfo  {publisher} {Cambridge University
  Press},\ \bibinfo {year} {2016})\ \Eprint {http://arxiv.org/abs/1212.5605}
  {arXiv:1212.5605 [hep-th]} \BibitemShut {NoStop}%
\bibitem [{\citenamefont {Arkani-Hamed}\ \emph
  {et~al.}(2017{\natexlab{a}})\citenamefont {Arkani-Hamed}, \citenamefont
  {Bai},\ and\ \citenamefont {Lam}}]{Arkani-Hamed:2017tmz}%
  \BibitemOpen
  \bibfield  {author} {\bibinfo {author} {\bibfnamefont {N.}~\bibnamefont
  {Arkani-Hamed}}, \bibinfo {author} {\bibfnamefont {Y.}~\bibnamefont {Bai}}, \
  and\ \bibinfo {author} {\bibfnamefont {T.}~\bibnamefont {Lam}},\ }\href
  {\doibase 10.1007/JHEP11(2017)039} {\bibfield  {journal} {\bibinfo  {journal}
  {JHEP}\ }\textbf {\bibinfo {volume} {11}},\ \bibinfo {pages} {039} (\bibinfo
  {year} {2017}{\natexlab{a}})},\ \Eprint {http://arxiv.org/abs/1703.04541}
  {arXiv:1703.04541 [hep-th]} \BibitemShut {NoStop}%
\bibitem [{\citenamefont {Lam}(2016)}]{Lam:2014jda}%
  \BibitemOpen
  \bibfield  {author} {\bibinfo {author} {\bibfnamefont {T.}~\bibnamefont
  {Lam}},\ }\href {\doibase 10.1007/s00220-016-2602-2} {\bibfield  {journal}
  {\bibinfo  {journal} {Commun. Math. Phys.}\ }\textbf {\bibinfo {volume}
  {343}},\ \bibinfo {pages} {1025} (\bibinfo {year} {2016})},\ \Eprint
  {http://arxiv.org/abs/1408.5531} {arXiv:1408.5531 [math.AG]} \BibitemShut
  {NoStop}%
\bibitem [{\citenamefont {Karp}\ and\ \citenamefont
  {Williams}(2017)}]{Karp_2017}%
  \BibitemOpen
  \bibfield  {author} {\bibinfo {author} {\bibfnamefont {S.~N.}\ \bibnamefont
  {Karp}}\ and\ \bibinfo {author} {\bibfnamefont {L.~K.}\ \bibnamefont
  {Williams}},\ }\href {\doibase 10.1093/imrn/rnx140} {\bibfield  {journal}
  {\bibinfo  {journal} {International Mathematics Research Notices}\ }\textbf
  {\bibinfo {volume} {2019}},\ \bibinfo {pages} {1401} (\bibinfo {year}
  {2017})}\BibitemShut {NoStop}%
\bibitem [{\citenamefont {Galashin}\ and\ \citenamefont
  {Lam}(2020)}]{Galashin_2020}%
  \BibitemOpen
  \bibfield  {author} {\bibinfo {author} {\bibfnamefont {P.}~\bibnamefont
  {Galashin}}\ and\ \bibinfo {author} {\bibfnamefont {T.}~\bibnamefont {Lam}},\
  }\href {\doibase 10.1112/s0010437x20007411} {\bibfield  {journal} {\bibinfo
  {journal} {Compositio Mathematica}\ }\textbf {\bibinfo {volume} {156}},\
  \bibinfo {pages} {2207} (\bibinfo {year} {2020})}\BibitemShut {NoStop}%
\bibitem [{\citenamefont {Franco}\ \emph {et~al.}(2015)\citenamefont {Franco},
  \citenamefont {Galloni}, \citenamefont {Mariotti},\ and\ \citenamefont
  {Trnka}}]{Franco:2014csa}%
  \BibitemOpen
  \bibfield  {author} {\bibinfo {author} {\bibfnamefont {S.}~\bibnamefont
  {Franco}}, \bibinfo {author} {\bibfnamefont {D.}~\bibnamefont {Galloni}},
  \bibinfo {author} {\bibfnamefont {A.}~\bibnamefont {Mariotti}}, \ and\
  \bibinfo {author} {\bibfnamefont {J.}~\bibnamefont {Trnka}},\ }\href
  {\doibase 10.1007/JHEP03(2015)128} {\bibfield  {journal} {\bibinfo  {journal}
  {JHEP}\ }\textbf {\bibinfo {volume} {03}},\ \bibinfo {pages} {128} (\bibinfo
  {year} {2015})},\ \Eprint {http://arxiv.org/abs/1408.3410} {arXiv:1408.3410
  [hep-th]} \BibitemShut {NoStop}%
\bibitem [{\citenamefont {Galloni}(2016)}]{Galloni:2016iuj}%
  \BibitemOpen
  \bibfield  {author} {\bibinfo {author} {\bibfnamefont {D.}~\bibnamefont
  {Galloni}},\ }\href@noop {} {\  (\bibinfo {year} {2016})},\ \Eprint
  {http://arxiv.org/abs/1601.02639} {arXiv:1601.02639 [hep-th]} \BibitemShut
  {NoStop}%
\bibitem [{\citenamefont {Bai}\ and\ \citenamefont {He}(2015)}]{Bai:2014cna}%
  \BibitemOpen
  \bibfield  {author} {\bibinfo {author} {\bibfnamefont {Y.}~\bibnamefont
  {Bai}}\ and\ \bibinfo {author} {\bibfnamefont {S.}~\bibnamefont {He}},\
  }\href {\doibase 10.1007/JHEP02(2015)065} {\bibfield  {journal} {\bibinfo
  {journal} {JHEP}\ }\textbf {\bibinfo {volume} {02}},\ \bibinfo {pages} {065}
  (\bibinfo {year} {2015})},\ \Eprint {http://arxiv.org/abs/1408.2459}
  {arXiv:1408.2459 [hep-th]} \BibitemShut {NoStop}%
\bibitem [{\citenamefont {Bai}\ \emph {et~al.}(2016)\citenamefont {Bai},
  \citenamefont {He},\ and\ \citenamefont {Lam}}]{Bai:2015qoa}%
  \BibitemOpen
  \bibfield  {author} {\bibinfo {author} {\bibfnamefont {Y.}~\bibnamefont
  {Bai}}, \bibinfo {author} {\bibfnamefont {S.}~\bibnamefont {He}}, \ and\
  \bibinfo {author} {\bibfnamefont {T.}~\bibnamefont {Lam}},\ }\href {\doibase
  10.1007/JHEP01(2016)112} {\bibfield  {journal} {\bibinfo  {journal} {JHEP}\
  }\textbf {\bibinfo {volume} {01}},\ \bibinfo {pages} {112} (\bibinfo {year}
  {2016})},\ \Eprint {http://arxiv.org/abs/1510.03553} {arXiv:1510.03553
  [hep-th]} \BibitemShut {NoStop}%
\bibitem [{\citenamefont {Rao}(2019)}]{Rao:2018uta}%
  \BibitemOpen
  \bibfield  {author} {\bibinfo {author} {\bibfnamefont {J.}~\bibnamefont
  {Rao}},\ }\href {\doibase 10.1016/j.nuclphysb.2019.114625} {\bibfield
  {journal} {\bibinfo  {journal} {Nucl. Phys. B}\ }\textbf {\bibinfo {volume}
  {943}},\ \bibinfo {pages} {114625} (\bibinfo {year} {2019})},\ \Eprint
  {http://arxiv.org/abs/1806.01765} {arXiv:1806.01765 [hep-th]} \BibitemShut
  {NoStop}%
\bibitem [{\citenamefont {Ferro}\ \emph {et~al.}(2019)\citenamefont {Ferro},
  \citenamefont {\L{}ukowski},\ and\ \citenamefont {Parisi}}]{Ferro:2018vpf}%
  \BibitemOpen
  \bibfield  {author} {\bibinfo {author} {\bibfnamefont {L.}~\bibnamefont
  {Ferro}}, \bibinfo {author} {\bibfnamefont {T.}~\bibnamefont {\L{}ukowski}},
  \ and\ \bibinfo {author} {\bibfnamefont {M.}~\bibnamefont {Parisi}},\ }\href
  {\doibase 10.1088/1751-8121/aaf3c3} {\bibfield  {journal} {\bibinfo
  {journal} {J. Phys. A}\ }\textbf {\bibinfo {volume} {52}},\ \bibinfo {pages}
  {045201} (\bibinfo {year} {2019})},\ \Eprint
  {http://arxiv.org/abs/1805.01301} {arXiv:1805.01301 [hep-th]} \BibitemShut
  {NoStop}%
\bibitem [{\citenamefont {Arkani-Hamed}\ \emph
  {et~al.}(2019{\natexlab{a}})\citenamefont {Arkani-Hamed}, \citenamefont
  {Langer}, \citenamefont {Yelleshpur~Srikant},\ and\ \citenamefont
  {Trnka}}]{Arkani-Hamed:2018rsk}%
  \BibitemOpen
  \bibfield  {author} {\bibinfo {author} {\bibfnamefont {N.}~\bibnamefont
  {Arkani-Hamed}}, \bibinfo {author} {\bibfnamefont {C.}~\bibnamefont
  {Langer}}, \bibinfo {author} {\bibfnamefont {A.}~\bibnamefont
  {Yelleshpur~Srikant}}, \ and\ \bibinfo {author} {\bibfnamefont
  {J.}~\bibnamefont {Trnka}},\ }\href {\doibase 10.1103/PhysRevLett.122.051601}
  {\bibfield  {journal} {\bibinfo  {journal} {Phys. Rev. Lett.}\ }\textbf
  {\bibinfo {volume} {122}},\ \bibinfo {pages} {051601} (\bibinfo {year}
  {2019}{\natexlab{a}})},\ \Eprint {http://arxiv.org/abs/1810.08208}
  {arXiv:1810.08208 [hep-th]} \BibitemShut {NoStop}%
\bibitem [{\citenamefont {Langer}\ and\ \citenamefont
  {Yelleshpur~Srikant}(2019)}]{Langer:2019iuo}%
  \BibitemOpen
  \bibfield  {author} {\bibinfo {author} {\bibfnamefont {C.}~\bibnamefont
  {Langer}}\ and\ \bibinfo {author} {\bibfnamefont {A.}~\bibnamefont
  {Yelleshpur~Srikant}},\ }\href {\doibase 10.1007/JHEP04(2019)105} {\bibfield
  {journal} {\bibinfo  {journal} {JHEP}\ }\textbf {\bibinfo {volume} {04}},\
  \bibinfo {pages} {105} (\bibinfo {year} {2019})},\ \Eprint
  {http://arxiv.org/abs/1902.05951} {arXiv:1902.05951 [hep-th]} \BibitemShut
  {NoStop}%
\bibitem [{\citenamefont {Bern}\ \emph {et~al.}(2007)\citenamefont {Bern},
  \citenamefont {Carrasco}, \citenamefont {Johansson},\ and\ \citenamefont
  {Kosower}}]{Bern:2007ct}%
  \BibitemOpen
  \bibfield  {author} {\bibinfo {author} {\bibfnamefont {Z.}~\bibnamefont
  {Bern}}, \bibinfo {author} {\bibfnamefont {J.~J.~M.}\ \bibnamefont
  {Carrasco}}, \bibinfo {author} {\bibfnamefont {H.}~\bibnamefont {Johansson}},
  \ and\ \bibinfo {author} {\bibfnamefont {D.~A.}\ \bibnamefont {Kosower}},\
  }\href {\doibase 10.1103/PhysRevD.76.125020} {\bibfield  {journal} {\bibinfo
  {journal} {Phys. Rev. D}\ }\textbf {\bibinfo {volume} {76}},\ \bibinfo
  {pages} {125020} (\bibinfo {year} {2007})},\ \Eprint
  {http://arxiv.org/abs/0705.1864} {arXiv:0705.1864 [hep-th]} \BibitemShut
  {NoStop}%
\bibitem [{\citenamefont {Bourjaily}\ \emph {et~al.}(2012)\citenamefont
  {Bourjaily}, \citenamefont {DiRe}, \citenamefont {Shaikh}, \citenamefont
  {Spradlin},\ and\ \citenamefont {Volovich}}]{Bourjaily:2011hi}%
  \BibitemOpen
  \bibfield  {author} {\bibinfo {author} {\bibfnamefont {J.~L.}\ \bibnamefont
  {Bourjaily}}, \bibinfo {author} {\bibfnamefont {A.}~\bibnamefont {DiRe}},
  \bibinfo {author} {\bibfnamefont {A.}~\bibnamefont {Shaikh}}, \bibinfo
  {author} {\bibfnamefont {M.}~\bibnamefont {Spradlin}}, \ and\ \bibinfo
  {author} {\bibfnamefont {A.}~\bibnamefont {Volovich}},\ }\href {\doibase
  10.1007/JHEP03(2012)032} {\bibfield  {journal} {\bibinfo  {journal} {JHEP}\
  }\textbf {\bibinfo {volume} {03}},\ \bibinfo {pages} {032} (\bibinfo {year}
  {2012})},\ \Eprint {http://arxiv.org/abs/1112.6432} {arXiv:1112.6432
  [hep-th]} \BibitemShut {NoStop}%
\bibitem [{\citenamefont {Bourjaily}\ \emph
  {et~al.}(2016{\natexlab{a}})\citenamefont {Bourjaily}, \citenamefont
  {Heslop},\ and\ \citenamefont {Tran}}]{Bourjaily:2015bpz}%
  \BibitemOpen
  \bibfield  {author} {\bibinfo {author} {\bibfnamefont {J.~L.}\ \bibnamefont
  {Bourjaily}}, \bibinfo {author} {\bibfnamefont {P.}~\bibnamefont {Heslop}}, \
  and\ \bibinfo {author} {\bibfnamefont {V.-V.}\ \bibnamefont {Tran}},\ }\href
  {\doibase 10.1103/PhysRevLett.116.191602} {\bibfield  {journal} {\bibinfo
  {journal} {Phys. Rev. Lett.}\ }\textbf {\bibinfo {volume} {116}},\ \bibinfo
  {pages} {191602} (\bibinfo {year} {2016}{\natexlab{a}})},\ \Eprint
  {http://arxiv.org/abs/1512.07912} {arXiv:1512.07912 [hep-th]} \BibitemShut
  {NoStop}%
\bibitem [{\citenamefont {Bourjaily}\ \emph
  {et~al.}(2016{\natexlab{b}})\citenamefont {Bourjaily}, \citenamefont
  {Heslop},\ and\ \citenamefont {Tran}}]{Bourjaily:2016evz}%
  \BibitemOpen
  \bibfield  {author} {\bibinfo {author} {\bibfnamefont {J.~L.}\ \bibnamefont
  {Bourjaily}}, \bibinfo {author} {\bibfnamefont {P.}~\bibnamefont {Heslop}}, \
  and\ \bibinfo {author} {\bibfnamefont {V.-V.}\ \bibnamefont {Tran}},\ }\href
  {\doibase 10.1007/JHEP11(2016)125} {\bibfield  {journal} {\bibinfo  {journal}
  {JHEP}\ }\textbf {\bibinfo {volume} {11}},\ \bibinfo {pages} {125} (\bibinfo
  {year} {2016}{\natexlab{b}})},\ \Eprint {http://arxiv.org/abs/1609.00007}
  {arXiv:1609.00007 [hep-th]} \BibitemShut {NoStop}%
\bibitem [{\citenamefont {Bern}\ \emph {et~al.}(2016)\citenamefont {Bern},
  \citenamefont {Herrmann}, \citenamefont {Litsey}, \citenamefont
  {Stankowicz},\ and\ \citenamefont {Trnka}}]{Bern:2015ple}%
  \BibitemOpen
  \bibfield  {author} {\bibinfo {author} {\bibfnamefont {Z.}~\bibnamefont
  {Bern}}, \bibinfo {author} {\bibfnamefont {E.}~\bibnamefont {Herrmann}},
  \bibinfo {author} {\bibfnamefont {S.}~\bibnamefont {Litsey}}, \bibinfo
  {author} {\bibfnamefont {J.}~\bibnamefont {Stankowicz}}, \ and\ \bibinfo
  {author} {\bibfnamefont {J.}~\bibnamefont {Trnka}},\ }\href {\doibase
  10.1007/JHEP06(2016)098} {\bibfield  {journal} {\bibinfo  {journal} {JHEP}\
  }\textbf {\bibinfo {volume} {06}},\ \bibinfo {pages} {098} (\bibinfo {year}
  {2016})},\ \Eprint {http://arxiv.org/abs/1512.08591} {arXiv:1512.08591
  [hep-th]} \BibitemShut {NoStop}%
\bibitem [{\citenamefont {Arkani-Hamed}\ \emph
  {et~al.}(2017{\natexlab{b}})\citenamefont {Arkani-Hamed}, \citenamefont
  {Benincasa},\ and\ \citenamefont {Postnikov}}]{Arkani-Hamed:2017fdk}%
  \BibitemOpen
  \bibfield  {author} {\bibinfo {author} {\bibfnamefont {N.}~\bibnamefont
  {Arkani-Hamed}}, \bibinfo {author} {\bibfnamefont {P.}~\bibnamefont
  {Benincasa}}, \ and\ \bibinfo {author} {\bibfnamefont {A.}~\bibnamefont
  {Postnikov}},\ }\href@noop {} {\  (\bibinfo {year} {2017}{\natexlab{b}})},\
  \Eprint {http://arxiv.org/abs/1709.02813} {arXiv:1709.02813 [hep-th]}
  \BibitemShut {NoStop}%
\bibitem [{\citenamefont {Arkani-Hamed}\ and\ \citenamefont
  {Benincasa}(2018)}]{Arkani-Hamed:2018bjr}%
  \BibitemOpen
  \bibfield  {author} {\bibinfo {author} {\bibfnamefont {N.}~\bibnamefont
  {Arkani-Hamed}}\ and\ \bibinfo {author} {\bibfnamefont {P.}~\bibnamefont
  {Benincasa}},\ }\href@noop {} {\  (\bibinfo {year} {2018})},\ \Eprint
  {http://arxiv.org/abs/1811.01125} {arXiv:1811.01125 [hep-th]} \BibitemShut
  {NoStop}%
\bibitem [{\citenamefont {Arkani-Hamed}\ \emph
  {et~al.}(2018{\natexlab{b}})\citenamefont {Arkani-Hamed}, \citenamefont
  {Bai}, \citenamefont {He},\ and\ \citenamefont {Yan}}]{Arkani-Hamed:2017mur}%
  \BibitemOpen
  \bibfield  {author} {\bibinfo {author} {\bibfnamefont {N.}~\bibnamefont
  {Arkani-Hamed}}, \bibinfo {author} {\bibfnamefont {Y.}~\bibnamefont {Bai}},
  \bibinfo {author} {\bibfnamefont {S.}~\bibnamefont {He}}, \ and\ \bibinfo
  {author} {\bibfnamefont {G.}~\bibnamefont {Yan}},\ }\href {\doibase
  10.1007/JHEP05(2018)096} {\bibfield  {journal} {\bibinfo  {journal} {JHEP}\
  }\textbf {\bibinfo {volume} {05}},\ \bibinfo {pages} {096} (\bibinfo {year}
  {2018}{\natexlab{b}})},\ \Eprint {http://arxiv.org/abs/1711.09102}
  {arXiv:1711.09102 [hep-th]} \BibitemShut {NoStop}%
\bibitem [{\citenamefont {Salvatori}(2018)}]{Salvatori:2018aha}%
  \BibitemOpen
  \bibfield  {author} {\bibinfo {author} {\bibfnamefont {G.}~\bibnamefont
  {Salvatori}},\ }\href@noop {} {\  (\bibinfo {year} {2018})},\ \Eprint
  {http://arxiv.org/abs/1806.01842} {arXiv:1806.01842 [hep-th]} \BibitemShut
  {NoStop}%
\bibitem [{\citenamefont {Arkani-Hamed}\ \emph
  {et~al.}(2019{\natexlab{b}})\citenamefont {Arkani-Hamed}, \citenamefont {He},
  \citenamefont {Salvatori},\ and\ \citenamefont
  {Thomas}}]{Arkani-Hamed:2019vag}%
  \BibitemOpen
  \bibfield  {author} {\bibinfo {author} {\bibfnamefont {N.}~\bibnamefont
  {Arkani-Hamed}}, \bibinfo {author} {\bibfnamefont {S.}~\bibnamefont {He}},
  \bibinfo {author} {\bibfnamefont {G.}~\bibnamefont {Salvatori}}, \ and\
  \bibinfo {author} {\bibfnamefont {H.}~\bibnamefont {Thomas}},\ }\href@noop {}
  {\  (\bibinfo {year} {2019}{\natexlab{b}})},\ \Eprint
  {http://arxiv.org/abs/1912.12948} {arXiv:1912.12948 [hep-th]} \BibitemShut
  {NoStop}%
\bibitem [{\citenamefont {Arkani-Hamed}\ \emph {et~al.}(2021)\citenamefont
  {Arkani-Hamed}, \citenamefont {He},\ and\ \citenamefont
  {Lam}}]{Arkani-Hamed:2019mrd}%
  \BibitemOpen
  \bibfield  {author} {\bibinfo {author} {\bibfnamefont {N.}~\bibnamefont
  {Arkani-Hamed}}, \bibinfo {author} {\bibfnamefont {S.}~\bibnamefont {He}}, \
  and\ \bibinfo {author} {\bibfnamefont {T.}~\bibnamefont {Lam}},\ }\href
  {\doibase 10.1007/JHEP02(2021)069} {\bibfield  {journal} {\bibinfo  {journal}
  {JHEP}\ }\textbf {\bibinfo {volume} {02}},\ \bibinfo {pages} {069} (\bibinfo
  {year} {2021})},\ \Eprint {http://arxiv.org/abs/1912.08707} {arXiv:1912.08707
  [hep-th]} \BibitemShut {NoStop}%
\bibitem [{\citenamefont {Arkani-Hamed}\ \emph
  {et~al.}(2019{\natexlab{c}})\citenamefont {Arkani-Hamed}, \citenamefont {He},
  \citenamefont {Lam},\ and\ \citenamefont {Thomas}}]{Arkani-Hamed:2019plo}%
  \BibitemOpen
  \bibfield  {author} {\bibinfo {author} {\bibfnamefont {N.}~\bibnamefont
  {Arkani-Hamed}}, \bibinfo {author} {\bibfnamefont {S.}~\bibnamefont {He}},
  \bibinfo {author} {\bibfnamefont {T.}~\bibnamefont {Lam}}, \ and\ \bibinfo
  {author} {\bibfnamefont {H.}~\bibnamefont {Thomas}},\ }\href@noop {} {\
  (\bibinfo {year} {2019}{\natexlab{c}})},\ \Eprint
  {http://arxiv.org/abs/1912.11764} {arXiv:1912.11764 [hep-th]} \BibitemShut
  {NoStop}%
\bibitem [{\citenamefont {He}\ and\ \citenamefont {Zhang}(2018)}]{He:2018okq}%
  \BibitemOpen
  \bibfield  {author} {\bibinfo {author} {\bibfnamefont {S.}~\bibnamefont
  {He}}\ and\ \bibinfo {author} {\bibfnamefont {C.}~\bibnamefont {Zhang}},\
  }\href {\doibase 10.1007/JHEP10(2018)054} {\bibfield  {journal} {\bibinfo
  {journal} {JHEP}\ }\textbf {\bibinfo {volume} {10}},\ \bibinfo {pages} {054}
  (\bibinfo {year} {2018})},\ \Eprint {http://arxiv.org/abs/1807.11051}
  {arXiv:1807.11051 [hep-th]} \BibitemShut {NoStop}%
\bibitem [{\citenamefont {Damgaard}\ \emph {et~al.}(2019)\citenamefont
  {Damgaard}, \citenamefont {Ferro}, \citenamefont {Lukowski},\ and\
  \citenamefont {Parisi}}]{Damgaard:2019ztj}%
  \BibitemOpen
  \bibfield  {author} {\bibinfo {author} {\bibfnamefont {D.}~\bibnamefont
  {Damgaard}}, \bibinfo {author} {\bibfnamefont {L.}~\bibnamefont {Ferro}},
  \bibinfo {author} {\bibfnamefont {T.}~\bibnamefont {Lukowski}}, \ and\
  \bibinfo {author} {\bibfnamefont {M.}~\bibnamefont {Parisi}},\ }\href
  {\doibase 10.1007/JHEP08(2019)042} {\bibfield  {journal} {\bibinfo  {journal}
  {JHEP}\ }\textbf {\bibinfo {volume} {08}},\ \bibinfo {pages} {042} (\bibinfo
  {year} {2019})},\ \Eprint {http://arxiv.org/abs/1905.04216} {arXiv:1905.04216
  [hep-th]} \BibitemShut {NoStop}%
\bibitem [{\citenamefont {Damgaard}\ \emph {et~al.}(2021)\citenamefont
  {Damgaard}, \citenamefont {Ferro}, \citenamefont {Lukowski},\ and\
  \citenamefont {Moerman}}]{Damgaard:2020eox}%
  \BibitemOpen
  \bibfield  {author} {\bibinfo {author} {\bibfnamefont {D.}~\bibnamefont
  {Damgaard}}, \bibinfo {author} {\bibfnamefont {L.}~\bibnamefont {Ferro}},
  \bibinfo {author} {\bibfnamefont {T.}~\bibnamefont {Lukowski}}, \ and\
  \bibinfo {author} {\bibfnamefont {R.}~\bibnamefont {Moerman}},\ }\href
  {\doibase 10.1007/JHEP02(2021)041} {\bibfield  {journal} {\bibinfo  {journal}
  {JHEP}\ }\textbf {\bibinfo {volume} {02}},\ \bibinfo {pages} {041} (\bibinfo
  {year} {2021})},\ \Eprint {http://arxiv.org/abs/2010.15858} {arXiv:2010.15858
  [hep-th]} \BibitemShut {NoStop}%
\bibitem [{\citenamefont {Aharony}\ \emph {et~al.}(2008)\citenamefont
  {Aharony}, \citenamefont {Bergman}, \citenamefont {Jafferis},\ and\
  \citenamefont {Maldacena}}]{Aharony:2008ug}%
  \BibitemOpen
  \bibfield  {author} {\bibinfo {author} {\bibfnamefont {O.}~\bibnamefont
  {Aharony}}, \bibinfo {author} {\bibfnamefont {O.}~\bibnamefont {Bergman}},
  \bibinfo {author} {\bibfnamefont {D.~L.}\ \bibnamefont {Jafferis}}, \ and\
  \bibinfo {author} {\bibfnamefont {J.}~\bibnamefont {Maldacena}},\ }\href
  {\doibase 10.1088/1126-6708/2008/10/091} {\bibfield  {journal} {\bibinfo
  {journal} {JHEP}\ }\textbf {\bibinfo {volume} {10}},\ \bibinfo {pages} {091}
  (\bibinfo {year} {2008})},\ \Eprint {http://arxiv.org/abs/0806.1218}
  {arXiv:0806.1218 [hep-th]} \BibitemShut {NoStop}%
\bibitem [{\citenamefont {Huang}\ and\ \citenamefont
  {Lipstein}(2010)}]{Huang:2010qy}%
  \BibitemOpen
  \bibfield  {author} {\bibinfo {author} {\bibfnamefont {Y.-t.}\ \bibnamefont
  {Huang}}\ and\ \bibinfo {author} {\bibfnamefont {A.~E.}\ \bibnamefont
  {Lipstein}},\ }\href {\doibase 10.1007/JHEP11(2010)076} {\bibfield  {journal}
  {\bibinfo  {journal} {JHEP}\ }\textbf {\bibinfo {volume} {11}},\ \bibinfo
  {pages} {076} (\bibinfo {year} {2010})},\ \Eprint
  {http://arxiv.org/abs/1008.0041} {arXiv:1008.0041 [hep-th]} \BibitemShut
  {NoStop}%
\bibitem [{\citenamefont {Gang}\ \emph {et~al.}(2011)\citenamefont {Gang},
  \citenamefont {Huang}, \citenamefont {Koh}, \citenamefont {Lee},\ and\
  \citenamefont {Lipstein}}]{Gang:2010gy}%
  \BibitemOpen
  \bibfield  {author} {\bibinfo {author} {\bibfnamefont {D.}~\bibnamefont
  {Gang}}, \bibinfo {author} {\bibfnamefont {Y.-t.}\ \bibnamefont {Huang}},
  \bibinfo {author} {\bibfnamefont {E.}~\bibnamefont {Koh}}, \bibinfo {author}
  {\bibfnamefont {S.}~\bibnamefont {Lee}}, \ and\ \bibinfo {author}
  {\bibfnamefont {A.~E.}\ \bibnamefont {Lipstein}},\ }\href {\doibase
  10.1007/JHEP03(2011)116} {\bibfield  {journal} {\bibinfo  {journal} {JHEP}\
  }\textbf {\bibinfo {volume} {03}},\ \bibinfo {pages} {116} (\bibinfo {year}
  {2011})},\ \Eprint {http://arxiv.org/abs/1012.5032} {arXiv:1012.5032
  [hep-th]} \BibitemShut {NoStop}%
\bibitem [{\citenamefont {Bianchi}\ \emph
  {et~al.}(2012{\natexlab{a}})\citenamefont {Bianchi}, \citenamefont {Leoni},
  \citenamefont {Mauri}, \citenamefont {Penati},\ and\ \citenamefont
  {Santambrogio}}]{Bianchi:2012cq}%
  \BibitemOpen
  \bibfield  {author} {\bibinfo {author} {\bibfnamefont {M.~S.}\ \bibnamefont
  {Bianchi}}, \bibinfo {author} {\bibfnamefont {M.}~\bibnamefont {Leoni}},
  \bibinfo {author} {\bibfnamefont {A.}~\bibnamefont {Mauri}}, \bibinfo
  {author} {\bibfnamefont {S.}~\bibnamefont {Penati}}, \ and\ \bibinfo {author}
  {\bibfnamefont {A.}~\bibnamefont {Santambrogio}},\ }\href {\doibase
  10.1007/JHEP07(2012)029} {\bibfield  {journal} {\bibinfo  {journal} {JHEP}\
  }\textbf {\bibinfo {volume} {07}},\ \bibinfo {pages} {029} (\bibinfo {year}
  {2012}{\natexlab{a}})},\ \Eprint {http://arxiv.org/abs/1204.4407}
  {arXiv:1204.4407 [hep-th]} \BibitemShut {NoStop}%
\bibitem [{\citenamefont {Brandhuber}\ \emph {et~al.}(2012)\citenamefont
  {Brandhuber}, \citenamefont {Travaglini},\ and\ \citenamefont
  {Wen}}]{Brandhuber:2012wy}%
  \BibitemOpen
  \bibfield  {author} {\bibinfo {author} {\bibfnamefont {A.}~\bibnamefont
  {Brandhuber}}, \bibinfo {author} {\bibfnamefont {G.}~\bibnamefont
  {Travaglini}}, \ and\ \bibinfo {author} {\bibfnamefont {C.}~\bibnamefont
  {Wen}},\ }\href {\doibase 10.1007/JHEP10(2012)145} {\bibfield  {journal}
  {\bibinfo  {journal} {JHEP}\ }\textbf {\bibinfo {volume} {10}},\ \bibinfo
  {pages} {145} (\bibinfo {year} {2012})},\ \Eprint
  {http://arxiv.org/abs/1207.6908} {arXiv:1207.6908 [hep-th]} \BibitemShut
  {NoStop}%
\bibitem [{\citenamefont {Bargheer}\ \emph {et~al.}(2012)\citenamefont
  {Bargheer}, \citenamefont {Beisert}, \citenamefont {Loebbert},\ and\
  \citenamefont {McLoughlin}}]{Bargheer:2012cp}%
  \BibitemOpen
  \bibfield  {author} {\bibinfo {author} {\bibfnamefont {T.}~\bibnamefont
  {Bargheer}}, \bibinfo {author} {\bibfnamefont {N.}~\bibnamefont {Beisert}},
  \bibinfo {author} {\bibfnamefont {F.}~\bibnamefont {Loebbert}}, \ and\
  \bibinfo {author} {\bibfnamefont {T.}~\bibnamefont {McLoughlin}},\ }\href
  {\doibase 10.1088/1751-8113/45/47/475402} {\bibfield  {journal} {\bibinfo
  {journal} {J. Phys. A}\ }\textbf {\bibinfo {volume} {45}},\ \bibinfo {pages}
  {475402} (\bibinfo {year} {2012})},\ \Eprint {http://arxiv.org/abs/1204.4406}
  {arXiv:1204.4406 [hep-th]} \BibitemShut {NoStop}%
\bibitem [{\citenamefont {Chen}\ and\ \citenamefont
  {Huang}(2011)}]{Chen:2011vv}%
  \BibitemOpen
  \bibfield  {author} {\bibinfo {author} {\bibfnamefont {W.-M.}\ \bibnamefont
  {Chen}}\ and\ \bibinfo {author} {\bibfnamefont {Y.-t.}\ \bibnamefont
  {Huang}},\ }\href {\doibase 10.1007/JHEP11(2011)057} {\bibfield  {journal}
  {\bibinfo  {journal} {JHEP}\ }\textbf {\bibinfo {volume} {11}},\ \bibinfo
  {pages} {057} (\bibinfo {year} {2011})},\ \Eprint
  {http://arxiv.org/abs/1107.2710} {arXiv:1107.2710 [hep-th]} \BibitemShut
  {NoStop}%
\bibitem [{\citenamefont {Bianchi}\ \emph
  {et~al.}(2012{\natexlab{b}})\citenamefont {Bianchi}, \citenamefont {Leoni},
  \citenamefont {Mauri}, \citenamefont {Penati},\ and\ \citenamefont
  {Santambrogio}}]{Bianchi:2011dg}%
  \BibitemOpen
  \bibfield  {author} {\bibinfo {author} {\bibfnamefont {M.~S.}\ \bibnamefont
  {Bianchi}}, \bibinfo {author} {\bibfnamefont {M.}~\bibnamefont {Leoni}},
  \bibinfo {author} {\bibfnamefont {A.}~\bibnamefont {Mauri}}, \bibinfo
  {author} {\bibfnamefont {S.}~\bibnamefont {Penati}}, \ and\ \bibinfo {author}
  {\bibfnamefont {A.}~\bibnamefont {Santambrogio}},\ }\href {\doibase
  10.1007/JHEP01(2012)056} {\bibfield  {journal} {\bibinfo  {journal} {JHEP}\
  }\textbf {\bibinfo {volume} {01}},\ \bibinfo {pages} {056} (\bibinfo {year}
  {2012}{\natexlab{b}})},\ \Eprint {http://arxiv.org/abs/1107.3139}
  {arXiv:1107.3139 [hep-th]} \BibitemShut {NoStop}%
\bibitem [{\citenamefont {Bianchi}\ and\ \citenamefont
  {Leoni}(2014)}]{Bianchi:2014iia}%
  \BibitemOpen
  \bibfield  {author} {\bibinfo {author} {\bibfnamefont {M.~S.}\ \bibnamefont
  {Bianchi}}\ and\ \bibinfo {author} {\bibfnamefont {M.}~\bibnamefont
  {Leoni}},\ }\href {\doibase 10.1007/JHEP11(2014)077} {\bibfield  {journal}
  {\bibinfo  {journal} {JHEP}\ }\textbf {\bibinfo {volume} {11}},\ \bibinfo
  {pages} {077} (\bibinfo {year} {2014})},\ \Eprint
  {http://arxiv.org/abs/1403.3398} {arXiv:1403.3398 [hep-th]} \BibitemShut
  {NoStop}%
\bibitem [{\citenamefont {Huang}\ \emph {et~al.}(2022)\citenamefont {Huang},
  \citenamefont {Kojima}, \citenamefont {Wen},\ and\ \citenamefont
  {Zhang}}]{Huang:2021jlh}%
  \BibitemOpen
  \bibfield  {author} {\bibinfo {author} {\bibfnamefont {Y.-t.}\ \bibnamefont
  {Huang}}, \bibinfo {author} {\bibfnamefont {R.}~\bibnamefont {Kojima}},
  \bibinfo {author} {\bibfnamefont {C.}~\bibnamefont {Wen}}, \ and\ \bibinfo
  {author} {\bibfnamefont {S.-Q.}\ \bibnamefont {Zhang}},\ }\href {\doibase
  10.1007/JHEP01(2022)141} {\bibfield  {journal} {\bibinfo  {journal} {JHEP}\
  }\textbf {\bibinfo {volume} {01}},\ \bibinfo {pages} {141} (\bibinfo {year}
  {2022})},\ \Eprint {http://arxiv.org/abs/2111.03037} {arXiv:2111.03037
  [hep-th]} \BibitemShut {NoStop}%
\bibitem [{\citenamefont {He}\ \emph {et~al.}(2022)\citenamefont {He},
  \citenamefont {Kuo},\ and\ \citenamefont {Zhang}}]{He:2021llb}%
  \BibitemOpen
  \bibfield  {author} {\bibinfo {author} {\bibfnamefont {S.}~\bibnamefont
  {He}}, \bibinfo {author} {\bibfnamefont {C.-K.}\ \bibnamefont {Kuo}}, \ and\
  \bibinfo {author} {\bibfnamefont {Y.-Q.}\ \bibnamefont {Zhang}},\ }\href
  {\doibase 10.1007/JHEP02(2022)148} {\bibfield  {journal} {\bibinfo  {journal}
  {JHEP}\ }\textbf {\bibinfo {volume} {02}},\ \bibinfo {pages} {148} (\bibinfo
  {year} {2022})},\ \Eprint {http://arxiv.org/abs/2111.02576} {arXiv:2111.02576
  [hep-th]} \BibitemShut {NoStop}%
\bibitem [{\citenamefont {Arkani-Hamed}\ \emph {et~al.}(2022)\citenamefont
  {Arkani-Hamed}, \citenamefont {Henn},\ and\ \citenamefont
  {Trnka}}]{Arkani-Hamed:2021iya}%
  \BibitemOpen
  \bibfield  {author} {\bibinfo {author} {\bibfnamefont {N.}~\bibnamefont
  {Arkani-Hamed}}, \bibinfo {author} {\bibfnamefont {J.}~\bibnamefont {Henn}},
  \ and\ \bibinfo {author} {\bibfnamefont {J.}~\bibnamefont {Trnka}},\ }\href
  {\doibase 10.1007/JHEP03(2022)108} {\bibfield  {journal} {\bibinfo  {journal}
  {JHEP}\ }\textbf {\bibinfo {volume} {03}},\ \bibinfo {pages} {108} (\bibinfo
  {year} {2022})},\ \Eprint {http://arxiv.org/abs/2112.06956} {arXiv:2112.06956
  [hep-th]} \BibitemShut {NoStop}%
\bibitem [{\citenamefont {Alday}\ \emph {et~al.}(2011)\citenamefont {Alday},
  \citenamefont {Buchbinder},\ and\ \citenamefont {Tseytlin}}]{Alday:2011ga}%
  \BibitemOpen
  \bibfield  {author} {\bibinfo {author} {\bibfnamefont {L.~F.}\ \bibnamefont
  {Alday}}, \bibinfo {author} {\bibfnamefont {E.~I.}\ \bibnamefont
  {Buchbinder}}, \ and\ \bibinfo {author} {\bibfnamefont {A.~A.}\ \bibnamefont
  {Tseytlin}},\ }\href {\doibase 10.1007/JHEP09(2011)034} {\bibfield  {journal}
  {\bibinfo  {journal} {JHEP}\ }\textbf {\bibinfo {volume} {09}},\ \bibinfo
  {pages} {034} (\bibinfo {year} {2011})},\ \Eprint
  {http://arxiv.org/abs/1107.5702} {arXiv:1107.5702 [hep-th]} \BibitemShut
  {NoStop}%
\bibitem [{\citenamefont {Engelund}\ and\ \citenamefont
  {Roiban}(2012)}]{Engelund:2011fg}%
  \BibitemOpen
  \bibfield  {author} {\bibinfo {author} {\bibfnamefont {O.~T.}\ \bibnamefont
  {Engelund}}\ and\ \bibinfo {author} {\bibfnamefont {R.}~\bibnamefont
  {Roiban}},\ }\href {\doibase 10.1007/JHEP05(2012)158} {\bibfield  {journal}
  {\bibinfo  {journal} {JHEP}\ }\textbf {\bibinfo {volume} {05}},\ \bibinfo
  {pages} {158} (\bibinfo {year} {2012})},\ \Eprint
  {http://arxiv.org/abs/1110.0758} {arXiv:1110.0758 [hep-th]} \BibitemShut
  {NoStop}%
\bibitem [{\citenamefont {Alday}\ \emph {et~al.}(2013)\citenamefont {Alday},
  \citenamefont {Henn},\ and\ \citenamefont {Sikorowski}}]{Alday:2013ip}%
  \BibitemOpen
  \bibfield  {author} {\bibinfo {author} {\bibfnamefont {L.~F.}\ \bibnamefont
  {Alday}}, \bibinfo {author} {\bibfnamefont {J.~M.}\ \bibnamefont {Henn}}, \
  and\ \bibinfo {author} {\bibfnamefont {J.}~\bibnamefont {Sikorowski}},\
  }\href {\doibase 10.1007/JHEP03(2013)058} {\bibfield  {journal} {\bibinfo
  {journal} {JHEP}\ }\textbf {\bibinfo {volume} {03}},\ \bibinfo {pages} {058}
  (\bibinfo {year} {2013})},\ \Eprint {http://arxiv.org/abs/1301.0149}
  {arXiv:1301.0149 [hep-th]} \BibitemShut {NoStop}%
\bibitem [{\citenamefont {Henn}\ \emph {et~al.}(2020)\citenamefont {Henn},
  \citenamefont {Korchemsky},\ and\ \citenamefont
  {Mistlberger}}]{Henn:2019swt}%
  \BibitemOpen
  \bibfield  {author} {\bibinfo {author} {\bibfnamefont {J.~M.}\ \bibnamefont
  {Henn}}, \bibinfo {author} {\bibfnamefont {G.~P.}\ \bibnamefont
  {Korchemsky}}, \ and\ \bibinfo {author} {\bibfnamefont {B.}~\bibnamefont
  {Mistlberger}},\ }\href {\doibase 10.1007/JHEP04(2020)018} {\bibfield
  {journal} {\bibinfo  {journal} {JHEP}\ }\textbf {\bibinfo {volume} {04}},\
  \bibinfo {pages} {018} (\bibinfo {year} {2020})},\ \Eprint
  {http://arxiv.org/abs/1911.10174} {arXiv:1911.10174 [hep-th]} \BibitemShut
  {NoStop}%
\bibitem [{\citenamefont {Chicherin}\ and\ \citenamefont
  {Henn}(2022{\natexlab{a}})}]{Chicherin:2022bov}%
  \BibitemOpen
  \bibfield  {author} {\bibinfo {author} {\bibfnamefont {D.}~\bibnamefont
  {Chicherin}}\ and\ \bibinfo {author} {\bibfnamefont {J.~M.}\ \bibnamefont
  {Henn}},\ }\href {\doibase 10.1007/JHEP07(2022)057} {\bibfield  {journal}
  {\bibinfo  {journal} {JHEP}\ }\textbf {\bibinfo {volume} {07}},\ \bibinfo
  {pages} {057} (\bibinfo {year} {2022}{\natexlab{a}})},\ \Eprint
  {http://arxiv.org/abs/2202.05596} {arXiv:2202.05596 [hep-th]} \BibitemShut
  {NoStop}%
\bibitem [{\citenamefont {Chicherin}\ and\ \citenamefont
  {Henn}(2022{\natexlab{b}})}]{Chicherin:2022zxo}%
  \BibitemOpen
  \bibfield  {author} {\bibinfo {author} {\bibfnamefont {D.}~\bibnamefont
  {Chicherin}}\ and\ \bibinfo {author} {\bibfnamefont {J.}~\bibnamefont
  {Henn}},\ }\href {\doibase 10.1007/JHEP07(2022)038} {\bibfield  {journal}
  {\bibinfo  {journal} {JHEP}\ }\textbf {\bibinfo {volume} {07}},\ \bibinfo
  {pages} {038} (\bibinfo {year} {2022}{\natexlab{b}})},\ \Eprint
  {http://arxiv.org/abs/2204.00329} {arXiv:2204.00329 [hep-th]} \BibitemShut
  {NoStop}%
\bibitem [{\citenamefont {Hodges}(2013)}]{Hodges:2009hk}%
  \BibitemOpen
  \bibfield  {author} {\bibinfo {author} {\bibfnamefont {A.}~\bibnamefont
  {Hodges}},\ }\href {\doibase 10.1007/JHEP05(2013)135} {\bibfield  {journal}
  {\bibinfo  {journal} {JHEP}\ }\textbf {\bibinfo {volume} {05}},\ \bibinfo
  {pages} {135} (\bibinfo {year} {2013})},\ \Eprint
  {http://arxiv.org/abs/0905.1473} {arXiv:0905.1473 [hep-th]} \BibitemShut
  {NoStop}%
\bibitem [{\citenamefont {Elvang}\ \emph {et~al.}(2014)\citenamefont {Elvang},
  \citenamefont {Huang}, \citenamefont {Keeler}, \citenamefont {Lam},
  \citenamefont {Olson}, \citenamefont {Roland},\ and\ \citenamefont
  {Speyer}}]{Elvang:2014fja}%
  \BibitemOpen
  \bibfield  {author} {\bibinfo {author} {\bibfnamefont {H.}~\bibnamefont
  {Elvang}}, \bibinfo {author} {\bibfnamefont {Y.-t.}\ \bibnamefont {Huang}},
  \bibinfo {author} {\bibfnamefont {C.}~\bibnamefont {Keeler}}, \bibinfo
  {author} {\bibfnamefont {T.}~\bibnamefont {Lam}}, \bibinfo {author}
  {\bibfnamefont {T.~M.}\ \bibnamefont {Olson}}, \bibinfo {author}
  {\bibfnamefont {S.~B.}\ \bibnamefont {Roland}}, \ and\ \bibinfo {author}
  {\bibfnamefont {D.~E.}\ \bibnamefont {Speyer}},\ }\href {\doibase
  10.1007/JHEP12(2014)181} {\bibfield  {journal} {\bibinfo  {journal} {JHEP}\
  }\textbf {\bibinfo {volume} {12}},\ \bibinfo {pages} {181} (\bibinfo {year}
  {2014})},\ \Eprint {http://arxiv.org/abs/1410.0621} {arXiv:1410.0621
  [hep-th]} \BibitemShut {NoStop}%
\bibitem [{Note1()}]{Note1}%
  \BibitemOpen
  \bibinfo {note} {See Supplemental Material for the detail of the reduction
  and the computation of their forms, which includes Ref.~\cite
  {stanley1973acyclic}.}\BibitemShut {Stop}%
\bibitem [{\citenamefont {Henn}\ \emph {et~al.}(2010)\citenamefont {Henn},
  \citenamefont {Plefka},\ and\ \citenamefont {Wiegandt}}]{Henn:2010ps}%
  \BibitemOpen
  \bibfield  {author} {\bibinfo {author} {\bibfnamefont {J.~M.}\ \bibnamefont
  {Henn}}, \bibinfo {author} {\bibfnamefont {J.}~\bibnamefont {Plefka}}, \ and\
  \bibinfo {author} {\bibfnamefont {K.}~\bibnamefont {Wiegandt}},\ }\href
  {\doibase 10.1007/JHEP11(2011)053} {\bibfield  {journal} {\bibinfo  {journal}
  {JHEP}\ }\textbf {\bibinfo {volume} {08}},\ \bibinfo {pages} {032} (\bibinfo
  {year} {2010})},\ \bibinfo {note} {[Erratum: JHEP 11, 053 (2011)]},\ \Eprint
  {http://arxiv.org/abs/1004.0226} {arXiv:1004.0226 [hep-th]} \BibitemShut
  {NoStop}%
\bibitem [{\citenamefont {Huang}\ and\ \citenamefont
  {Wen}(2014)}]{Huang:2013owa}%
  \BibitemOpen
  \bibfield  {author} {\bibinfo {author} {\bibfnamefont {Y.-T.}\ \bibnamefont
  {Huang}}\ and\ \bibinfo {author} {\bibfnamefont {C.}~\bibnamefont {Wen}},\
  }\href {\doibase 10.1007/JHEP02(2014)104} {\bibfield  {journal} {\bibinfo
  {journal} {JHEP}\ }\textbf {\bibinfo {volume} {02}},\ \bibinfo {pages} {104}
  (\bibinfo {year} {2014})},\ \Eprint {http://arxiv.org/abs/1309.3252}
  {arXiv:1309.3252 [hep-th]} \BibitemShut {NoStop}%
\bibitem [{\citenamefont {Huang}\ \emph {et~al.}(2014)\citenamefont {Huang},
  \citenamefont {Wen},\ and\ \citenamefont {Xie}}]{Huang:2014xza}%
  \BibitemOpen
  \bibfield  {author} {\bibinfo {author} {\bibfnamefont {Y.-t.}\ \bibnamefont
  {Huang}}, \bibinfo {author} {\bibfnamefont {C.}~\bibnamefont {Wen}}, \ and\
  \bibinfo {author} {\bibfnamefont {D.}~\bibnamefont {Xie}},\ }\href {\doibase
  10.1088/1751-8113/47/47/474008} {\bibfield  {journal} {\bibinfo  {journal}
  {J. Phys.}\ }\textbf {\bibinfo {volume} {A47}},\ \bibinfo {pages} {474008}
  (\bibinfo {year} {2014})},\ \Eprint {http://arxiv.org/abs/1402.1479}
  {arXiv:1402.1479 [hep-th]} \BibitemShut {NoStop}%
\bibitem [{\citenamefont {Caron-Huot}\ and\ \citenamefont
  {Huang}(2013)}]{Caron-Huot:2012sos}%
  \BibitemOpen
  \bibfield  {author} {\bibinfo {author} {\bibfnamefont {S.}~\bibnamefont
  {Caron-Huot}}\ and\ \bibinfo {author} {\bibfnamefont {Y.-t.}\ \bibnamefont
  {Huang}},\ }\href {\doibase 10.1007/JHEP03(2013)075} {\bibfield  {journal}
  {\bibinfo  {journal} {JHEP}\ }\textbf {\bibinfo {volume} {03}},\ \bibinfo
  {pages} {075} (\bibinfo {year} {2013})},\ \Eprint
  {http://arxiv.org/abs/1210.4226} {arXiv:1210.4226 [hep-th]} \BibitemShut
  {NoStop}%
\bibitem [{\citenamefont {Stanley}(1973)}]{stanley1973acyclic}%
  \BibitemOpen
  \bibfield  {author} {\bibinfo {author} {\bibfnamefont {R.~P.}\ \bibnamefont
  {Stanley}},\ }\href@noop {} {\bibfield  {journal} {\bibinfo  {journal}
  {Discrete Mathematics}\ }\textbf {\bibinfo {volume} {5}},\ \bibinfo {pages}
  {171} (\bibinfo {year} {1973})}\BibitemShut {NoStop}%
\bibitem [{Note2()}]{Note2}%
  \BibitemOpen
  \bibinfo {note} {Note that naively we could have spurious boundaries with
  $y_j'=y_i'$ or $w_j'=w_i'$, but they are forbidden by the mutual conditions
  ${z'}_{i,j}^2-y'_{i,j} w'_{i,j}<0$, thus only physical poles can appear for
  each bipartite geometry.}\BibitemShut {Stop}%
\end{thebibliography}%

\newpage

\widetext
\begin{center}
\textbf{\Large Supplementary Material}
\end{center}

\section{Details of transitive reduction and bipartite geometries}


In this section, we study ${\cal A}_L$ and associated negative geometries~\cite{Arkani-Hamed:2021iya}, and show that they are much simpler than the $D=4$ case while having very rich structures. We will see that the decomposition of the reduced amplituhedron into negative geometries is highly redundant, and a much more efficient way is to formulate ${\cal A}_L$ directly in terms of bipartite graphs, which we call {\it bipartite geometries}.\\

A nice trick for studying the $n=4$ amplituhedron~\cite{Arkani-Hamed:2013kca} is to rewrite its canonical form $\Omega_L$ as the sum with alternating signs of canonical forms of negative geometries represented by all possible undirected graphs with $L$ nodes without $2$-cycles, 
\begin{equation}
\Omega_L=\sum_{\text{graph $g$}}(-1)^{E(g)}\Omega_g,
\end{equation}
where $E(g)$ is the number of edges of the graph $g$, and $\Omega_g$ is the canonical form of the negative geometry represented by the graph $g$ which is defined by $\forall \text{node } i: x_i, y_i, z_i, w_i>0$ and mutual negative conditions:
\begin{equation}
\forall \text{ edge $i \text{ --- } j$}: x_{i,j} z_{i,j} + y_{i,j} w_{i,j}>0 .
\end{equation}

It suffices to consider all {\it connected} graphs, whose sum with alternating signs gives the geometry for the logarithm of amplitudes~\cite{Arkani-Hamed:2021iya}. For example, $\Omega_3$ is given by the sum of $\Omega_1^3/3!$, $-\tilde{\Omega}_2 \Omega_1/2$, and the connected part $\tilde{\Omega}_3$ (with chain and triangle graphs):
\begin{equation}
    \includegraphics{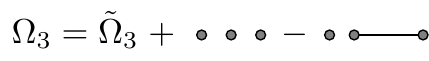},\quad 
    \includegraphics{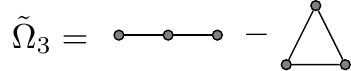}.
\end{equation}
Similarly, the connected part of $L=4$ is given by the sum of graphs with $6$ topologies (and so on for higher $L$),
\begin{equation}
    \includegraphics{figs/Omega4t.pdf}
\end{equation}

What is new in $D=3$ is that we will see most of these geometries do not contribute at all! We find that the negative geometries are directly related to time-ordered regions for points in AdS${}_3$; this then leads to the surprising conclusion that only those associated with {\it bipartite} graphs survive, with very simple pole structures. 

Already in $D=4$, the mutual condition $x_{i,j} z_{i,j} + y_{i,j} w_{i,j}$ can be viewed as the distance square between two points in $\mathbb{R}^{2,2}$, once we notice {\it e.g.} $(x, z)=\phi^2\pm \phi^3$, $(y, w):=\phi^0 \pm \phi^1$. In $D=3$, the additional symplectic condition $xz+yw=1$ means $\phi_i$ further live on the unit sphere in $\mathbb R^{2,2}$, which is in fact the AdS$_3$ by definition. The AdS$_3$ is further conformally flat to $\mathbb{R}^{1,2}$: after solving for $x=(1-yw)/z$, we make a change of variables $\Phi=(y'=y/z, w'=w/z,z'=1/z)$, and the distance square becomes 
\begin{equation}
x_{i,j}z_{i,j}+y_{i,j}w_{i,j}=\frac 1 {z'_i z'_j} (y'_{i,j} w'_{i,j}-{{z}_{i,j}'^2})=:\frac 1 {z'_i z'_j} (\Phi_i-\Phi_j)^2,
\end{equation}
which is essentially the metric of AdS$_3$ in Poincar\'e coordinates, where $\Phi=(\Phi^0,\Phi^1,\Phi^2):=(\frac{y'+w'}{2},\frac{y'-w'}{2},z')\in \mathbb R^{1,2}$.

Now we observe that two loops are connected by an edge, {\it i.e.} with mutual negativity conditions $x_{i,j} z_{i,j} + y_{i,j} w_{i,j}>0$, if and only if the two points in AdS$_3$ are {\bf time-like} separated
\begin{equation}
(\Phi_i-\Phi_j)^2=y'_{i,j} w'_{i,j}-{{z}_{i,j}'^2}>0.
\end{equation}
Positivity conditions for each loop read $y'>0,w'>0,z'>0$, as well as $x=({z'}^2-y' w')/z'=-\Phi^2/z'>0$ (or $\Phi^2<0$): the last condition implies that it is {\bf space-like} separated from the special point $(y', z', w')=(0,0,0)\equiv {\bf 0}$.

This simple observation has important implications for negative geometries. When two points are time-like separated, there is a natural time-ordering $\prec$ for loops such that $i \prec j$ if $\Phi_j$ is in the forward lightcone of $\Phi_i$. Then by adding an arrow $i\to j$ if $i\prec j$ for each edge $i \text{ --- } j$ in the graph, this means that any connected negative geometry is the sum of {\bf time-ordered regions} represented by {\it acyclic orientations} of the graph since any closed loop of time orderings leads to contradictions, {\it e.g.} $1\prec 2\prec 3$ but $3 \prec 1$ is impossible. In this way, we write ${\cal A}_L$ as the sum of all {\it directed acyclic graphs} (DAGs)~\cite{stanley1973acyclic}, weighted by $(-)^E$, and $\tilde{\Omega}_L$ is given by the canonical forms for those with connected DAGs. It is a classic problem for enumerating all acyclic orientations for a given graph. For a complete graph, we have exactly $L!$ such regions, one for each ordering of $L$ points; for other graphs, there are less than only the latter contribute to the sum $L!$ regions since some points are not ordered; in the other extreme, any tree diagram has $2^{L{-}1}$ regions. As the first nontrivial example, for $L=3$, we have $6+4\times 3=18$ DAGs: $6$ for the triangle and $4$ for each $3$ chain graph.


A huge reduction happens here due to the simple fact that some directed edges (mutual negativity conditions) may be useless in a DAG. If there are $i\to j$ and $j\to k$, we can see that these two edges imply $i\to k$ from the lightcone picture, so we can delete this edge (if it exists) while the geometry leaves unchanged. Therefore, for a DAG $G$ we try to find a minimal sub-diagram $H$ such that all edges of $G$ can be implied by edges in $H$, then $\mathcal A_G=\mathcal A_H$. This minimal sub-diagram $H$ is called a \textit{transitive reduction} of $G$, a well-known concept in graph theory. For a finite DAG, there exists a unique transitive reduction, which can be easily found by the depth-first search. 
Transitive reductions lead to the following theorem.

\vspace{1ex}

\noindent {\bf Theorem} Under transitive reductions, we have
\begin{equation}\label{eq:DAGBP}
\sum_{{\rm DAG}~G} (-)^{E(G)} {\cal A}_G=\sum_{{\rm bipartite~DAG}~g} (-)^{E(g)} {\cal A}_g,
\end{equation}
where all graphs are connected, and $E(G)$ is the number of edges in the graph $G$.

\vspace{1ex}

Bipartite DAGs on RHS are DAGs that only have sources or sinks, so we want to prove that the other DAGs all cancel in the summation on LHS.

We define $G\sim H$ if transitive reductions of $G$ and $H$ are the same, then $\mathcal A_G=\mathcal A_H$. This relation is an equivalence relation that divides all DAGs into disjoint equivalence classes. In fact, if two classes are not disjoint, any DAG in the intersection will have different transitive reductions, which contradicts the uniqueness of the transitive reduction. 

From the minimal diagram $H$ in an equivalence class, suppose its edges can imply $m$ extra edges, then we can see that there are $2^m$ diagrams in this class, $\binom{m}{k}$ for adding $k$ edges into $H$. Therefore, if $m>0$, the summation of all DAGs in this equivalence class is 
\begin{equation}
\sum_{\text {DAG } G\sim H}(-1)^{E(G)} \mathcal A_G =
\mathcal A_H\sum_{\text{DAG } G\sim H}(-1)^{E(G)}=
\mathcal A_H(-1)^{E_H}\sum_{k=0}^m(-1)^{k}{m \choose k}= \mathcal A_H (-1)^{E_H} (1-1)^m=0.
\end{equation}
The only left graphs are those with $m=0$, which are exactly graphs with only sources and sinks! In fact, if there is a node which is not a source nor a sink, there exists a sub-diagram $i\to j \to k$ where we can add $i\to k$, so $m>0$.

Therefore, we have proven the theorem. We use $L=4$ as an example. There are 10 topological inequivalent minimal DAGs. They are
\begin{equation}
\begin{aligned}
&m=0:\quad \includegraphics[align=c]{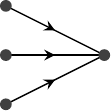}
\quad 
\includegraphics[align=c]{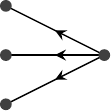}
\quad 
\includegraphics[align=c]{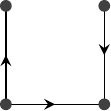}
\quad 
\includegraphics[align=c]{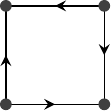},
\\
&m=1:\quad \includegraphics[align=c]{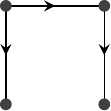}
\quad 
\includegraphics[align=c]{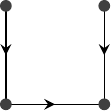}
\quad 
\includegraphics[align=c]{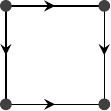},
\\
&m=2:\quad \includegraphics[align=c]{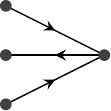}
\quad 
\includegraphics[align=c]{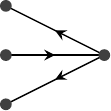},\\
&m=3:\quad \includegraphics[align=c]{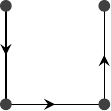},
\end{aligned}
\end{equation}
where we group them into different $m$. For each DAG, there are $2^m$ DAGs in the corresponding equivalence class. For example, the equivalence class of the first DAG with $m=2$ in the above table contains four DAGs
\begin{equation}\biggl\{\includegraphics[align=c]{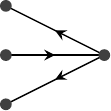}
\,,\,\includegraphics[align=c]{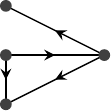}
\,,\,\includegraphics[align=c]{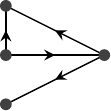}
\,,\,\includegraphics[align=c]{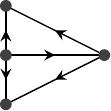}
\biggr\}
\end{equation}
and it's easy to see that
\begin{equation}\includegraphics[align=c]{figs/c1.pdf}
\,-\,\includegraphics[align=c]{figs/c2.pdf}
\,-\,\includegraphics[align=c]{figs/c3.pdf}
\,+\,\includegraphics[align=c]{figs/c4.pdf}
\,=\,0.
\end{equation}

The counting of all equivalence classes for general $L$ is still unknown, but here we list the counting $2, 12, 146, 3060, 101642, 5106612, \dots$ from $L=2$, which can also be found as A001927 at {\tt oeis.org} and understood as the number of connected partially ordered sets with $L$ labeled points. If we are not interested in relabels, there are $1, 3, 10, 44, 238, 1650, 14512, 163341, \dots$ topologically inequivalent classes for $L=2,3,\dots$, which is A000608 at {\tt oeis.org}. 

\section{Canonical forms of negative geometries for reduced amplituhedron}
In this section, we study canonical forms of negative geometries for the $n=4$ reduced amplituhedron in $D=3$. We will do so for geometries associated with general, non-bipartite DAGs as well as bipartite geometries (those associated with bipartite graphs). Even though we have shown that only the latter contributes to the sum of negative geometries, we do not restrict our discussion to bipartite graphs for now. We will see how each piece of geometries is reduced and simplified with dimension reduction, even before any cancellations happen in the sum. The most representative example is the $L$-loop complete graph, which is the most difficult one to compute in $D=4$, but it simply vanishes in $D=3$ for $L>2$. We will then give explicit examples of canonical forms for tree graphs and present a general formula for any trees. 

\subsection{From pole structures to vanishing forms}
We begin with analyzing the boundary structure of negative geometries. Recall that for any two loops, $x_{i,j} z_{i,j} + y_{i,j} w_{i,j}$ ($y'_{i,j} w'_{i,j}-{{z}_{i,j}'^2}$) is given by distance in AdS${}_3$, and we first show that for any two points which are time ordered, {\it e.g.} $i\prec j$, $x_i=0$, $z_i=0$, $y_j=0$, $w_j=0$ can not be the codimension-one boundaries of the geometries. Consequently, if a vertex is neither a source nor sink, then the vertex can only have $D_{i,j}=0$ (for $j$'s connected to $i$) as its pole. \\

Since $y'_j>y'_i>0$ (and similarly for $w'$), $j$ cannot have a pole with $y_j=0$ or $w_j=0$. On the other hand, since $j$ is space-like separated from ${\bf 0}$, {\it i.e.} ${z_j'}^2-y_j' w_j'>0$, so is $i$ (otherwise it leads to contradiction with time ordering), thus ${z_i'}^2-y_i' w_i' \propto x_i z_i =0$ cannot be a codimension-one boundary for $i$, thus $x_i=0$ or $z_i=0$ cannot be a pole~\footnote{Note that naively we could have spurious boundaries with $y_j'=y_i'$ or $w_j'=w_i'$, but they are forbidden by the mutual conditions ${z'}_{i,j}^2-y'_{i,j} w'_{i,j}<0$, thus only physical poles can appear for each bipartite geometry.}.\\

For bipartite graphs, we can associate a source/sink vertex (black/white node), which is ``earlier/later" than all points connected to it. It can only have $s \propto y w$ pole, or $t\propto x z$ pole, respectively. For the non-bipartite graphs (DAGs), we need to add the rule for a vertex that is neither source nor sink that such kind of vertex cannot have any pole of its own. 
\begin{equation}
    \includegraphics[scale=0.8]{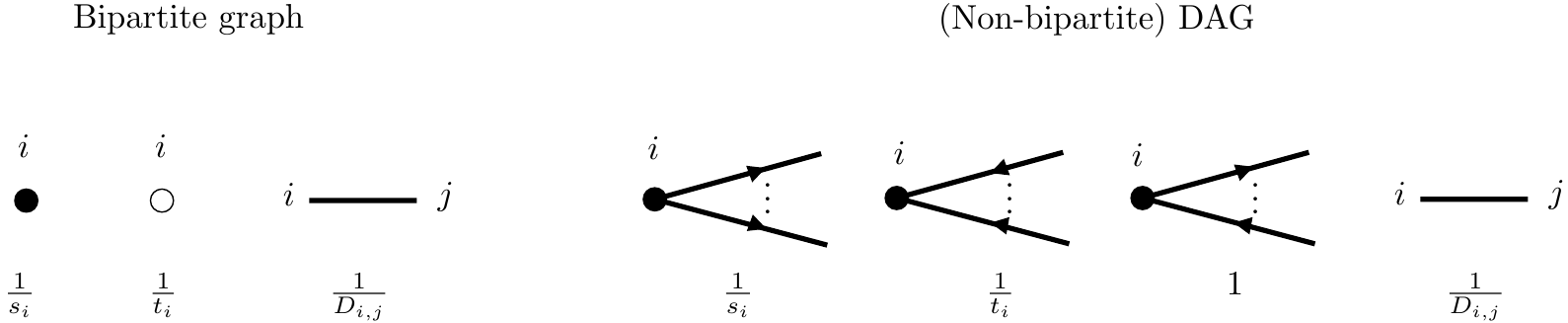}.
\end{equation}

Now that we have determined all possible poles of the form for each graph, we can use various constraints to fix the numerator. Let us first recall (DCI) weight {\it etc.} of the form: requiring dual conformal invariance (DCI) means that the form is invariant under individual rescaling of any loop variable $\ell_i$. The form with this property is called DCI weight neutral: for each loop $\ell_i$, its measure, $d^3 \ell_i$ has weighs 3, which means that the total weight in $\ell_i$ of its denominator minus that of the numerator needs to be 3. A pole $s_i$ or $t_i$ in the denominator has weight 2 in $\ell_i$ and the pole $D_{i,j}$ has weight 1 in $\ell_i$ and $\ell_j$. Our strategy is to write down all possible numerators with the correct weight as well as other symmetries, such as the symmetry under flipping the sign of $\epsilon_i$'s, the symmetry under the exchange $x_i \leftrightarrow z_i$ and $y_i \leftrightarrow w_i$, as well as graph symmetries of negative geometries. 

Before proceeding, we have seen that the transitive reduction and pole structures discussed above dramatically reduced the space of possible numerators for any bipartite graph: the former simplifies possible $D_{i,j}$ poles, and the latter halves the weight for each vertex, allowing only $s$ pole or $t$ pole. In the following, we will focus on some cases of non-bipartite graphs where the constraints are so strong  that it is impossible to write down any numerator for their form, thus the canonical forms vanish identically.\\

First, let us consider a special case: a non-bipartite DAGs with a valency-2 vertex that is neither source nor sink, then the form must vanish. From our rule of poles, if vertex $i$ is neither source or sink, the form cannot have $s_i$ or $t_i$ poles, but only mutual poles $D_{i,j}=0$ (for $j$'s connected to $i$). But since it has valency $2$, there are only $2$ such poles, thus the form in $\ell_i$, which has dimension 3, must vanish. Said differently, we cannot write down any numerator such that $\ell_i$ has the correct weight since the denominator has weight $2$. Thus it must vanish.\\

This simple fact implies that many DAGs (in particular, all complete graphs) vanish, as long as the graph has this kind of valency-2 vertex. For example, let us consider a complete graph with $D_{i,j}<0,\forall i,j=1,2,\dots,L$, {\it e.g.} triangle at $L=3$ and tetrahedron at $L=4$. There are $L!$ DAGs for all possible orderings of the $L$ points, and all of them are equivalent. Let us consider the region with ordering  $1\prec2\prec\dots\prec L$; by transitive reduction, all edges $(ij)$ with $|j-i|>1$ can be deleted, resulting in a chain graph where $L{-}2$ points, $2,3, \cdots, L{-}1$, have valency $2$. Any of them is neither {\color{red} a} source nor {\color{red} a} sink. Thus any such DAG vanishes (with respect to any of these $L{-}2$ loops). We should also notice that although at $L=3$, the non-vanishing DAGs are already bipartite graphs, this is no longer true starting from $L=4$. It is still crucial to sum over all DAGs with $(-)^E$ to cancel non-bipartite graphs. 

\subsection{Examples and general formula for canonical forms of tree graphs}
Now we move to canonical forms associated with bipartite graphs, and in particular we first give the forms for all tree graphs, and leave those for loop graphs to the next section. Recall that our method is to construct an ansatz of the numerator and then fix it by various constraints; it turns out that for any tree graph, the form can be completely solved in this way, which leads to a recursive construction for any tree form which can be proved using {\color{red} the} constraint of lower boundaries of geometry. The upshot is that, when attaching node $j$ to $i$ to construct a $L$-node tree, where node $i$ has valency $v_i$ in the original $(L{-}1)$-node tree, all we need is an ``inverse-soft factor" ${\cal T}_{j\to i}$:
\begin{equation}
    \includegraphics[align=c]{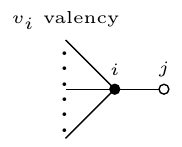}=
    \begin{cases}
    &\frac{2\epsilon_i}{D_{i,j}t_j},\quad  v_i \text{ odd }\\
    &\frac{2c t_i}{\epsilon_iD_{i,j}t_j}, \quad v_i \text{ even}
    \end{cases}\quad,\qquad
    \includegraphics[align=c]{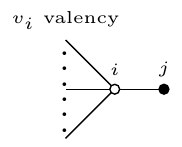}=
    \begin{cases}
    &\frac{2\epsilon_i}{D_{i,j}s_j}, \quad v_i \text{ odd}\\
    &\frac{2c s_i}{\epsilon_iD_{i,j}s_j}, \quad v_i \text{ even}
    \end{cases}\quad.
\end{equation}

\noindent
According to the rule, the $L$-node tree form can be written as the $(L{-}1)$-node tree form times  ${\cal T}_{j\to i}$:
\begin{equation}
\underline{\Omega}_L^{\rm tree} (j \to i)=\underline{\Omega}_{L{-}1} ^{\rm tree}\times {\cal T}_{j \to i}.
\end{equation}
Forms for all trees are determined in this way.\\

Let us illustrate this recursive construction of tree forms with an explicit example. We consider the following $L=6$ tree graph which can be constructed recursively:
\begin{equation}
    \includegraphics[scale=0.7]{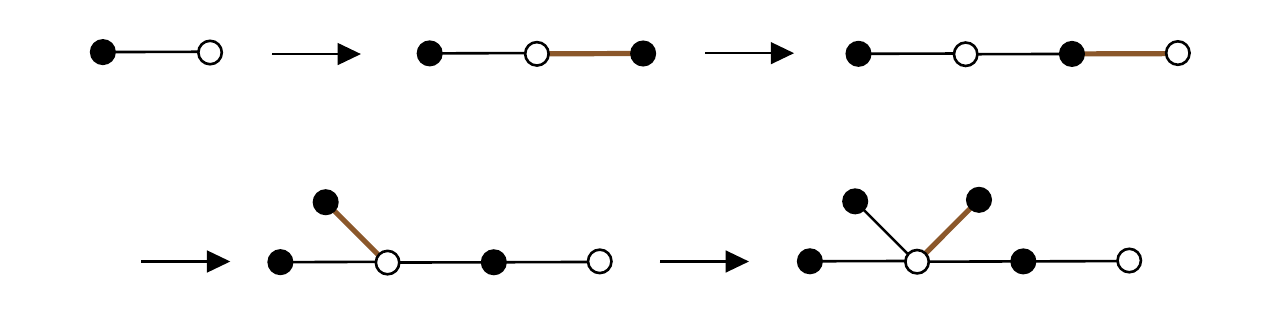}.
\end{equation}
\noindent
It begins with two nodes with one link connecting them. Based on the pole structure of the bipartite graph, its denominator is $s_1 t_1 D_{1,2}$; since each loop in the denominator has weight 3, its numerator can only be a constant, which is fixed to be $2c^2$ by requiring the form has the unit residue. In step two and three, we attach a black (white) node to a valence-one vertex, and the inverse soft factor ${\cal T}_{3\to 2}$ (${\cal T}_{4\to 3}$) is $2\epsilon_2/s_3D_{2,3}$ ($2\epsilon_3/t_4D_{3,4}$). In step four, we attach a black node to $\ell_2$ (valence-two vertex)  which introduces the factor $2c s_2/\epsilon_2s_5 D_{2,5}$. The  $\epsilon_2$ in the denominator of ${\cal{T}}_{5\to 2}$ is cancelled by the one in the numerator and does not introduce any new pole. In the final step, a black node is attached to t$\ell_2$ (valence-three vertex now), and the form is multiplied by a factor $2\epsilon_2/s_6 D_{2,6}$. We summarize how the form is constructed as follows:
\begin{equation}
    \begin{split}
        &\frac{2c^2}{s_1 t_2 D_{1,2}} \longrightarrow \frac{2c^2}{s_1 t_2 D_{1,2}} \times \frac{2 \epsilon_2}{s_3 D_{2,3}}\longrightarrow \frac{4c^{2}\epsilon_2}{s_1 t_2 s_3 D_{1,2}D_{2,3}} \times \frac{2 \epsilon_3}{t_4D_{3,4}}\\
        \longrightarrow &\frac{8c^{2}\epsilon_2 \epsilon_3}{s_1 t_2 s_3 t_4 D_{1,2}D_{2,3}D_{3,4}}\times \frac{2 c s_2}{\epsilon_2 s_5 D_{2,5}}\longrightarrow \frac{16 c^{3} \epsilon_3 s_2}{s_1 t_2 s_3 t_4 s_5 D_{1,2}D_{2,3}D_{2,5}D_{3,4}} \times \frac{2 \epsilon_2}{s_6 D_{2,6}}.
    \end{split}
\end{equation}
\noindent
The construction is path-independent. We obtain the same result along different paths leading to the same graph.\\

The recursive construction for tree graphs in $D=3$ depends on the valence and type (black or white) of the vertices, making it very different from the one in $D=4$. We can see that the $D=4$ form of the tree graphs factorizes to a product of two-loop building blocks~\cite{Arkani-Hamed:2021iya}:
\begin{equation}
    \begin{split}
        \tilde{\Omega}^{4D}_L=\prod_{k=1}^L \frac{d^4AB_k}{s_k t_k}\times \prod_{g}N_{i,j}, \quad \text{where} \quad N_{i,j}=-\frac{\langle\ell_i 13\rangle\langle\ell_j 24\rangle+\langle\ell_i 24\rangle\langle\ell_j 13\rangle}{\langle1234\rangle\langle\ell_i \ell_j\rangle},
    \end{split}
\end{equation}
where $N_{i,j}$ is the form two-loop negative link module the $s$ and $t$ poles. We have seen that {\color{red} the} $D=3$ case is quite different since there are two kinds of ``factorizations", and the non-locality of inverse-soft factors makes each factorization not independent. For example, the non-local pole in step four is canceled by the numerator in step two. 

Finally, the recursive construction leads to a closed form for general tree graphs. We denote the set of sources and sinks as $B$ and $W$ (for $L$ even, we have $|B|=|W|=L/2$ and for $L$ odd, $|B|=(L{+}1)/2$, $|W|=(L{-}1)/2$ or vice versa), then we can write 
\begin{equation} \label{eq:gentree}
\underline{\Omega}^{\rm tree}({\cal A}_g)=\frac{2^{L{-}1} c^{L/2+1} N_g}{\prod_{i \in B(g)} s_i \prod_{j \in W(g)} t_j \prod_{e \in E(g)} D_e},
\end{equation}
where $2^{L{-}1} c^{L/2+1}$ is needed for unit residue, and it is easy to see that the weight of the denominator for loop $i$ is $v_i+2$ with $v_i$ the valency, thus we need $N_g$ to have weight $v_i-1$ for $i$. One can show that for $i\in B$ with odd $v_i$, all we need is a factor $t_i^{(v_i{-}1)/2}$, and for even $v_i$, $t_i^{v_i/2-1} \epsilon_i/c^{1/2}$; the same for $j\in W$ with $s_j$ instead.

\section{Computation of $L=4,5$ integrands and consistency checks}
In this section, we compute canonical forms for remaining bipartite geometries which constitute loop integrands of $L=4,5$. Since we have solved the forms for tree graphs, all we need are forms for loop bipartite graphs, and we present details of the calculation for these graphs at $L=4,5$. Furthermore, we provide consistent checks for the resulting loop integrands.
 
\subsection{Example of loop bipartite graphs: box graph for $L=4$}
We begin with the calculation for the form of $L=4$ box graph:
\begin{equation}
\begin{split}
    \includegraphics[scale=0.7]{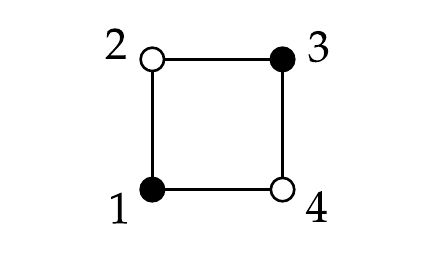}
\end{split}
\end{equation}
\noindent
According to the pole structure discussed above, it has $s_1 t_2 s_3 t_4 D_{1,2} D_{2,3} D_{3,4} D_{1,4}$ in the denominator. Since the denominator has weight $4$ for each loop, the numerator must have weight $1$ in $\ell_i$ ($i=1,2,3,4$). We find six types of terms that give the correct DCI weight:
\begin{align} 
    \{&\prod_{i=1}^{4}\epsilon_{i}, \epsilon_{i_{1}}\epsilon_{i_{2}}\langle \ell_{i_{3}}p_{i_{3}}q_{i_{3}}\rangle\langle \ell_{i_{4}}p_{i_{4}}q_{i_{4}}\rangle\langle1234\rangle,\epsilon_{i_1}\epsilon_{i_2}\langle \ell_{i_{3}}\ell_{i_{4}}\rangle\langle1234\rangle^2\nonumber\\
    &\prod_{i=1}^{4}\langle \ell_{i}p_{i}q_{i}\rangle\langle1234\rangle^{2},\langle \ell_{i_{1}}\ell_{i_{2}}\rangle\langle \ell_{i_{3}}p_{i_{3}}q_{i_{3}}\rangle\langle \ell_{i_{4}}p_{i_{4}}q_{i_{4}}\rangle\langle1234\rangle^3,\langle \ell_{i_{1}}\ell_{i_{2}}\rangle\langle \ell_{i_{3}}\ell_{i_{4}}\rangle\langle1234\rangle^{4}\},
\end{align}
where $p,q=1,2,3,4$ are external legs with correct weight. Note that we have only included terms with {\color{red} an} even number of $\epsilon$ factors, {\it i.e.} $0,2,4$, which reflects the fact that  $L=4$ integrand should be invariant under $\epsilon \to -\epsilon$. 
For $n=4$, with $\epsilon_i \to -\epsilon_i$ for $i=1,2,3,4$, the integrand picks up a sign factor $(-)^L$, which can be derived from the property of integrands under reflection symmetry~\cite{Bargheer:2012cp}. 
After imposing symmetries of exchanging $w_i\leftrightarrow y_i$ and $x_i\leftrightarrow z_i$ independently, as well as the symmetry between $\ell_1, \ell_3$ and $\ell_2, \ell_4$ of the graph, we find the total numbers of parameters in the ansatz is reduced to $25$.\\

Now, we use the lower boundaries to constrain the numerator. We call this type of boundaries $L$-cuts by cutting $y_i=0$ or $w_i=0$ for black vertices and $x_i=0$ or $z_i=0$ for white vertices, which can be calculated by direct triangulation. 
Such $L$-cuts simplify mutual conditions $D_{i,j}>0$ a lot: for example, $x_i=y_j=0$ (here we solve $y_i=(1-x_i z_i)/w_i$ and $x_j=(1-w_j y_j)/z_j$) implies
\begin{equation}
    \begin{split}
        D_{i,j}=2-\frac{w_j}{w_i}-\frac{z_i}{z_j}>0,
    \end{split}
\end{equation}
which makes direct triangulation much easier.\\

We present the following $L$-cut results for the form of the box:
\begin{itemize}
    \item $y_1=y_3=x_2=x_4=0$
    \begin{equation}
    -\frac{4 \left(w_1 w_3 z_1 z_3-4 w_2 w_4 z_1 z_3+w_2 w_4 z_2 z_4\right)}{w_1 w_3 z_2 z_4 \left.\left(D_{1,2}D_{2,3}D_{3,4}D_{1,4}\right)\right|_{x_2=x_4=y_1=y_3=0}}.
\end{equation}
\item $w_1=y_3=x_2=x_4=0$
\begin{equation}
    -\frac{4 \left(w_3 y_1 z_1 z_3-4 z_1 z_3+z_2 z_4\right)}{w_3 y_1 z_2 z_4 \left.\left(D_{1,2}D_{2,3}D_{3,4}D_{1,4}\right)\right|_{x_2=x_4=y_3=w_1=0}}.
\end{equation}
\item $y_1=y_3=z_2=x_4=0$
    \begin{equation}
    -\frac{4 \left(w_2 w_4 x_2 z_4+w_1 w_3-4 w_2 w_4\right)}{w_1 w_3 x_2 z_4 \left.\left(D_{1,2}D_{2,3}D_{3,4}D_{1,4}\right)\right|_{x_4=z_2=y_1=y_3=0}}.
\end{equation}
\item $w_1=y_3=z_2=x_4=0$
\begin{equation}
    -\frac{4 \left(w_3 y_1+x_2 z_4-4\right)}{w_3 x_2 y_1 z_4 \left.\left(D_{1,2}D_{2,3}D_{3,4}D_{1,4}\right)\right|_{x_4=z_2=w_1=y_3=0}}.
\end{equation}
\end{itemize}
These conditions suffice to fix all parameters in the ansatz, and we have  computed a large number of other cuts, providing numerous consistency checks for our calculation. \\

The same method can be applied to forms of other bipartite geometries for $L>4$. In addition to tree graphs, there are two topologies of loop graphs, $T_5$ and $T_6$ for $L=5$. Based on DCI weight, dihedral symmetry and graph symmetries, we find 60 parameters in the numerator of {\color{red} the} $T_5$ graph, and 172 parameters in that of the $T_6$ graph. In these cases, we need to impose more conditions compared to the box at $L=4$, but after imposing such conditions we indeed fix all parameters and arrive at the unique canonical form of $T_5$ and $T_6$, which have been presented in the main text.

\subsection{Consistency checks on ABJM loop integrands}
Finally, we give more details regarding what we have checked about the claim that the canonical form of reduced amplituhedron gives ABJM four-point integrands. What we have computed up to $L=5$ automatically satisfy those all-loop cuts discussed above, which have been essentially trivialized by our geometries. Such cuts are very powerful, {\it e.g.} in the context of generalized unitarity, soft cuts can determine coefficients of a large portion of dual conformal integrals. However, from {\color{red} the} geometry point of view, soft cuts can only probe disconnected graphs and cannot constrain those negative geometries contributing to $\tilde\Omega_L$~\cite{Arkani-Hamed:2021iya}; moreover, all vanishing cuts become manifest in our geometries as well. Therefore, we should consider more general cuts to which all topologies of graphs (including connected and disconnected ones) contribute. The double cuts, which follow from unitarity, seem to be perfect for that. 

Before discussing double cuts, we remark that there are also many other simple cuts, such as ladder and next-to-ladder cuts, which can be derived easily from geometries. However, due to a lack of data for $L\geq 4$ from the physics side, we do not use them as independent checks; we did confirm that such ladder-type cuts of our $L=3$ result, which agrees with the conjecture of~\cite{Bianchi:2014iia}, are satisfied by cutting those contributing Feynman diagrams. 

The double cut, or unitarity cut, is a particular cut on a single loop:
\begin{equation}
	\langle AB12\rangle=\langle AB34 \rangle=0 \quad \text{or equivalently} \quad w_1=y_1=0.
\end{equation}
From {\color{red} the} optical theorem, the residue is given in terms of products of lower-loop amplitudes with momentum shifted:
\begin{equation}\label{eq: double cut}
	\begin{split}
		\sum_{k=0}^{L-1}\frac{\vec{x}_{1,3}^2\, \langle l_1 p_1 \rangle \langle p_1 p_4 \rangle \langle p_4 l_1\rangle}{\vec{x}_{a,2}^2 \vec{x}_{a,4}^2} M_4^{k-\text{loop}}(\vec{x}_1,\vec{x}_2,\vec{x}_3,\vec{x}_a) M_4^{(L{-}k{-}1)-\text{loop}}(\vec{x}_1,\vec{x}_a,\vec{x}_3,\vec{x}_4)
	\end{split}
\end{equation} 
 
\begin{figure}[H]
    \centering
    \includegraphics{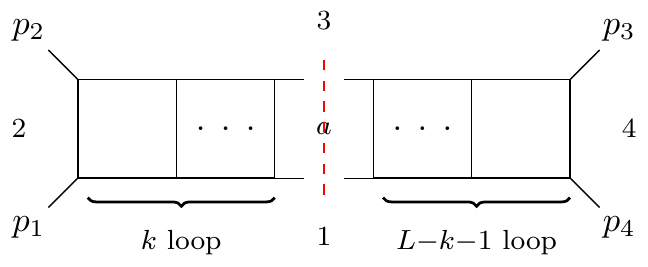}.
\end{figure}
Here, $\vec{x}_i$ is the dual variable and $\vec{x}_{i+1}-\vec{x}_i \equiv \vec{p_i}$.  The dual momentum variables of the left amplitude are $\vec{x}_1, \vec{x}_2, \vec{x}_3, \vec{x}_a$ and the dual variables of the right amplitude are $\vec{x}_1, \vec{x}_a, \vec{x}_3, \vec{x}_4$. The overall prefactor becomes unity when expressed in terms of $w, x, y, z$ variables.

To be more explicit, let us first consider the $L=3$ case. After cutting out $w_1=y_1=0$, the amplitude factorize to $A^{\text{tree}}_4\otimes A^{\text{2-loop}}_4$, $A^{\text{1-loop}}_4 \otimes A^{\text{1-loop}}_4$, and  $A^{\text{2-loop}}_4 \otimes A^{\text{tree}}_4$. Here the momentum is shifted and the $\epsilon$-numerator is shifted to $c^{3/2}(z_1-z_i)$ on the left side and to $c^{3/2}((-1+w_i y_i)z_1+z_i)/z_1 z_i$ on the right side, which are denoted as $\epsilon_{i,L}$ and $\epsilon_{i,R}$ respectively. Also $t_i$ is shifted to $c^2 D_{1,i}z_i$ on the left and $c^2 D_{1,i}x_i$ on the right, denoted as $t_{i,L}$ and $t_{i,R}$ respectively.  We find that the residue of $L=3$ integrand at $w_1=y_1=0$ nicely matches the sum of the three terms:
\begin{equation}
	\begin{split}
	& \text{Double cut}\  w_1=y_1=0\  \text{of the 3-loop integrand}\\
	    =& 1\times \left(\frac{1}{2}\frac{c^2\,\epsilon_{2,R}\epsilon_{3,R}}{s_{2}t_{2,R}s_{3}t_{3,R}}-\frac{2c^2x_{1}}{s_{2}t_{3,R}D_{2,3}}\right)+\frac{c\,\epsilon_{2,L}}{s_2 t_{2,L}}\times \frac{c\,\epsilon_{2,R}}{s_2 t_{2,R}}+\left(\frac{1}{2}\frac{\epsilon_{2,L}\epsilon_{3,L}}{s_{2}t_{2,L}s_{3}t_{3,L}}-\frac{2c^2z_{1}}{s_{2}t_{3,L}D_{2,3}}\right)\times 1+(2\leftrightarrow3)
	\end{split}.
\end{equation}

 For higher $L$, we proceed in exactly the same way: we first confirm that the $L=4$ form (after including all disconnected graphs) has the correct double cut, which uses the shifted amplitude of one- to three- loops; then we use the four-loop result and shift it to express the double cut of the $L=5$ form, which again has the correct double cut. This has provided solid evidence that the $L\leq 5$ results obtained from geometry are indeed correct!

\end{document}